\documentclass[preprint,final,5p,times,twocolumn]{elsarticle}
\usepackage{rotating,color,subfigure,amssymb}
\usepackage{alphalph}
\usepackage{amsmath}
\usepackage[T1]{fontenc}

\usepackage{amssymb}

\journal{Physics Letters B}
\begin{document}

\begin{frontmatter}



\title{Isoscalar Single-Pion Production in the Region of Roper and $d^*(2380)$ Resonances}

\author[IKPUU]{The WASA-at-COSY Collaboration\\[2ex] P.~Adlarson\fnref{fnmz}}
\author[ASWarsN]{W.~Augustyniak}
\author[IPJ]{W.~Bardan}
\author[Edinb]{M.~Bashkanov\corref{coau}}\ead{mikhail.bashkanov@ed.ac.uk}
\author[MS]{F.S.~Bergmann}
\author[ASWarsH]{M.~Ber{\l}owski}
\author[IITB]{H.~Bhatt}
\author[Budker,Novosib]{A.~Bondar}
\author[IKPJ]{M.~B\"uscher\fnref{fnpgi,fndus}}
\author[IKPUU]{H.~Cal\'{e}n}
\author[IFJ]{I.~Ciepa{\l}}
\author[PITue,Kepler]{H.~Clement}
\author[IPJ]{E.~Czerwi{\'n}ski}
\author[MS]{K.~Demmich}
\author[IKPJ]{R.~Engels}
\author[ZELJ]{A.~Erven}
\author[ZELJ]{W.~Erven}
\author[Erl]{W.~Eyrich}
\author[IKPJ,ITEP]{P.~Fedorets}
\author[Giess]{K.~F\"ohl}
\author[IKPUU]{K.~Fransson}
\author[IKPJ]{F.~Goldenbaum}
\author[IKPJ,IITI]{A.~Goswami}
\author[IKPJ,HepGat]{K.~Grigoryev\fnref{fnac}}
\author[IKPUU]{C.--O.~Gullstr\"om}
\author[IKPUU]{L.~Heijkenskj\"old\fnref{fnmz}}
\author[IKPJ]{V.~Hejny}
\author[MS]{N.~H\"usken}
\author[IPJ]{L.~Jarczyk}
\author[IKPUU]{T.~Johansson}
\author[IPJ]{B.~Kamys}
\author[ZELJ]{G.~Kemmerling\fnref{fnjcns}}
\author[IPJ]{G.~Khatri\fnref{fnharv}}
\author[MS]{A.~Khoukaz}
\author[IPJ]{O.~Khreptak}
\author[HiJINR]{D.A.~Kirillov}
\author[IPJ]{S.~Kistryn}
\author[ZELJ]{H.~Kleines\fnref{fnjcns}}
\author[Katow]{B.~K{\l}os}
\author[IPJ]{W.~Krzemie{\'n}}
\author[IFJ]{P.~Kulessa}
\author[IKPUU,ASWarsH]{A.~Kup{\'s}{\'c}}
\author[Budker,Novosib]{A.~Kuzmin}
\author[NITJ]{K.~Lalwani}
\author[IKPJ]{D.~Lersch}
\author[IKPJ]{B.~Lorentz}
\author[IPJ]{A.~Magiera}
\author[IKPJ,JARA]{R.~Maier}
\author[IKPUU]{P.~Marciniewski}
\author[ASWarsN]{B.~Maria{\'n}ski}
\author[ASWarsN]{H.--P.~Morsch}
\author[IPJ]{P.~Moskal}
\author[IKPJ]{H.~Ohm}
\author[IFJ]{W.~Parol}
\author[PITue,Kepler]{E.~Perez del Rio\fnref{fnlnf}}
\author[HiJINR]{N.M.~Piskunov}
\author[IKPJ]{D.~Prasuhn}
\author[IKPUU,ASWarsH]{D.~Pszczel}
\author[IFJ]{K.~Pysz}
\author[IKPUU,IPJ]{A.~Pyszniak}
\author[IKPJ,JARA,Bochum]{J.~Ritman}
\author[IITI]{A.~Roy}
\author[IPJ]{Z.~Rudy}
\author[IPJ]{O.~Rundel}
\author[IITB,IKPJ]{S.~Sawant}
\author[IKPJ]{S.~Schadmand}
\author[IPJ]{I.~Sch\"atti--Ozerianska}
\author[IKPJ]{T.~Sefzick}
\author[IKPJ]{V.~Serdyuk}
\author[Budker,Novosib]{B.~Shwartz}
\author[MS]{K.~Sitterberg}
\author[PITue,Kepler,Tomsk]{T.~Skorodko}
\author[IPJ]{M.~Skurzok}
\author[IPJ]{J.~Smyrski}
\author[ITEP]{V.~Sopov}
\author[IKPJ]{R.~Stassen}
\author[ASWarsH]{J.~Stepaniak}
\author[Katow]{E.~Stephan}
\author[IKPJ]{G.~Sterzenbach}
\author[IKPJ]{H.~Stockhorst}
\author[IKPJ,JARA]{H.~Str\"oher}
\author[IFJ]{A.~Szczurek}
\author[ASWarsN]{A.~Trzci{\'n}ski}
\author[IITB]{R.~Varma}
\author[IKPUU]{M.~Wolke}
\author[IPJ]{A.~Wro{\'n}ska}
\author[ZELJ]{P.~W\"ustner}
\author[KEK]{A.~Yamamoto}
\author[ASLodz]{J.~Zabierowski}
\author[IPJ]{M.J.~Zieli{\'n}ski}
\author[IKPUU]{J.~Z{\l}oma{\'n}czuk}
\author[ASWarsN]{P.~{\.Z}upra{\'n}ski}
\author[IKPJ]{M.~{\.Z}urek}

\address[IKPUU]{Division of Nuclear Physics, Department of Physics and 
 Astronomy, Uppsala University, Box 516, 75120 Uppsala, Sweden}
\address[ASWarsN]{Department of Nuclear Physics, National Centre for Nuclear 
 Research, ul.\ Hoza~69, 00-681, Warsaw, Poland}
\address[IPJ]{Institute of Physics, Jagiellonian University, prof.\ 
 Stanis{\l}awa {\L}ojasiewicza~11, 30-348 Krak\'{o}w, Poland}
\address[Edinb]{School of Physics and Astronomy, University of Edinburgh, 
 James Clerk Maxwell Building, Peter Guthrie Tait Road, Edinburgh EH9 3FD, 
 Great Britain}
\address[MS]{Institut f\"ur Kernphysik, Westf\"alische Wilhelms--Universit\"at 
 M\"unster, Wilhelm--Klemm--Str.~9, 48149 M\"unster, Germany}
\address[ASWarsH]{High Energy Physics Department, National Centre for Nuclear 
 Research, ul.\ Hoza~69, 00-681, Warsaw, Poland}
\address[IITB]{Department of Physics, Indian Institute of Technology Bombay, 
 Powai, Mumbai--400076, Maharashtra, India}
\address[Budker]{Budker Institute of Nuclear Physics of SB RAS, 11~akademika 
 Lavrentieva prospect, Novosibirsk, 630090, Russia}
\address[Novosib]{Novosibirsk State University, 2~Pirogova Str., Novosibirsk, 
 630090, Russia}
\address[IKPJ]{Institut f\"ur Kernphysik, Forschungszentrum J\"ulich, 52425 
 J\"ulich, Germany}
\address[IFJ]{The Henryk Niewodnicza{\'n}ski Institute of Nuclear Physics, 
 Polish Academy of Sciences, 152~Radzikowskiego St, 31-342 Krak\'{o}w, Poland}
\address[PITue]{Physikalisches Institut, Eberhard--Karls--Universit\"at 
 T\"ubingen, Auf der Morgenstelle~14, 72076 T\"ubingen, Germany}
\address[Kepler]{Kepler Center for Astro and Particle Physics, Eberhard Karls 
 University T\"ubingen, Auf der Morgenstelle~14, 72076 T\"ubingen, Germany}
\address[ZELJ]{Zentralinstitut f\"ur Engineering, Elektronik und Analytik, 
 Forschungszentrum J\"ulich, 52425 J\"ulich, Germany}
\address[Erl]{Physikalisches Institut, Friedrich--Alexander--Universit\"at 
 Erlangen--N\"urnberg, Erwin--Rommel-Str.~1, 91058 Erlangen, Germany}
\address[ITEP]{Institute for Theoretical and Experimental Physics, State 
 Scientific Center of the Russian Federation, Bolshaya Cheremushkinskaya~25, 
 117218 Moscow, Russia}
\address[Giess]{II.\ Physikalisches Institut, Justus--Liebig--Universit\"at 
 Gie{\ss}en, Heinrich--Buff--Ring~16, 35392 Giessen, Germany}
\address[IITI]{Department of Physics, Indian Institute of Technology Indore, 
 Khandwa Road, Indore--452017, Madhya Pradesh, India}
\address[HepGat]{High Energy Physics Division, Petersburg Nuclear Physics 
 Institute, Orlova Rosha~2, Gatchina, Leningrad district 188300, Russia}
\address[HiJINR]{Veksler and Baldin Laboratory of High Energiy Physics, Joint 
 Institute for Nuclear Physics, Joliot--Curie~6, 141980 Dubna, Moscow region, 
 Russia}
\address[Katow]{August Che{\l}kowski Institute of Physics, University of 
 Silesia, Uniwersytecka~4, 40-007, Katowice, Poland}
\address[NITJ]{Department of Physics, Malaviya National Institute of 
 Technology Jaipur, 302017, Rajasthan, India}
\address[JARA]{JARA--FAME, J\"ulich Aachen Research Alliance, Forschungszentrum 
 J\"ulich, 52425 J\"ulich, and RWTH Aachen, 52056 Aachen, Germany}
\address[Bochum]{Institut f\"ur Experimentalphysik I, Ruhr--Universit\"at 
 Bochum, Universit\"atsstr.~150, 44780 Bochum, Germany}
\address[Tomsk]{Department of Physics, Tomsk State University, 36~Lenina 
 Avenue, Tomsk, 634050, Russia}
\address[KEK]{High Energy Accelerator Research Organisation KEK, Tsukuba, 
 Ibaraki 305--0801, Japan}
\address[ASLodz]{Department of Astrophysics, National Centre for Nuclear 
 Research, Box~447, 90--950 {\L}\'{o}d\'{z}, Poland}
\address[HISKP]{Helmholtz--Institut f\"ur Strahlen-- und Kernphysik, 
 Rheinische Friedrich--Wilhelms--Universit\"at Bonn, Nu{\ss}allee~14--16, 
 53115 Bonn, Germany}

\fntext[fnmz]{present address: Institut f\"ur Kernphysik, Johannes 
 Gutenberg--Universit\"at Mainz, Johann--Joachim--Becher Weg~45, 55128 Mainz, 
 Germany}
\fntext[fnpgi]{present address: Peter Gr\"unberg Institut, PGI--6 Elektronische 
 Eigenschaften, Forschungszentrum J\"ulich, 52425 J\"ulich, Germany}
\fntext[fndus]{present address: Institut f\"ur Laser-- und Plasmaphysik, 
 Heinrich--Heine Universit\"at D\"usseldorf, Universit\"atsstr.~1, 40225 
 D\"usseldorf, Germany}
\fntext[fnac]{present address: III.~Physikalisches Institut~B, Physikzentrum, 
 RWTH Aachen, 52056 Aachen, Germany}
\fntext[fnjcns]{present address: J\"ulich Centre for Neutron Science JCNS, 
 Forschungszentrum J\"ulich, 52425 J\"ulich, Germany}
\fntext[fnharv]{present address: Department of Physics, Harvard University, 
 17~Oxford St., Cambridge, MA~02138, USA}
\fntext[fnlnf]{present address: INFN, Laboratori Nazionali di Frascati, Via 
 E.~Fermi, 40, 00044 Frascati (Roma), Italy}

\cortext[coau]{Corresponding author }


\begin{abstract}

Exclusive measurements of the quasi-free $pn \to pp\pi^-$  and $pp \to
pp\pi^0$ reactions have
been performed by means of $pd$ collisions at $T_p$ = 1.2 GeV using the WASA
detector setup at COSY. Total and differential cross sections have been
obtained covering the energy region $T_p = 0.95 - 1.3$ GeV ($\sqrt s$ = 2.3 -
2.46 GeV), which includes the regions of $\Delta(1232)$, $N^*(1440)$ and
$d^*(2380)$ resonance excitations. From these measurements the isoscalar
single-pion production has been extracted, for
which data existed so far only below $T_p$ = 1 GeV. We observe a substantial
increase of this cross section around 1 GeV, which can be related to the Roper
resonance $N^*(1440)$, the strength of which shows up isolated from the
$\Delta$ resonance in the isoscalar $(N\pi)_{I=0}$ invariant-mass spectrum. No
evidence for a decay of the dibaryon resonance $d^*(2380)$ into the isoscalar
$(NN\pi)_{I=0}$ channel is found. An upper limit of 
90 
$\mu$b (90
$\%$ C.L.) corresponding to a branching ratio of 
5 
$\%$ has been deduced.

\end{abstract}

\begin{keyword}
Single-Pion Production, Isoscalar Part, Roper Resonance, Dibaryon Resonance

\end{keyword}
\end{frontmatter}





\section{Introduction}

Single-pion production in nucleon-nucleon ($NN$) collisions may be separated
into isoscalar and isovector production. Excitation of the $\Delta(1232)$
resonance and of higher-lying $\Delta$ states in the course of the collision
process can only happen in an isovector process. Hence these are absent in
isoscalar single-pion production, which comprises only isoscalar processes
like the excitation of the Roper resonance 
$N^*(1440)$ and higher-lying $N^*$ resonances -- but also to the excitation and
decay of the recently observed dibaryon state $d^*(2380)$ with $I(J^P) = 0(3^+)$
\cite{MB,prl2011,np,npfull}

At incident energies below 1 GeV, single-pion production is strongly
characterized by excitation and decay of the $\Delta(1232)$ resonance. There
have been several attempts in the past to extract the isoscalar 
production cross section
\cite{Dakhno,Bys,Rappenecker,Thomas,Gatchina,Gatchina1}, in order to reveal
production processes other than 
the dominating $\Delta$ process.  Since single-pion production in
$NN$ collisions is either purely isovector or isospin-mixed, the isoscalar
cross section has to be obtained by combination of various cross section
measurements. Most often the relation \cite{Dakhno,Bys}:
\begin{eqnarray}
\sigma_{NN \to NN\pi}(I=0) = 3(2\sigma_{np \to pp\pi^-} - \sigma_{pp \to
  pp\pi^0})
\end{eqnarray}
is used. Since here the difference of two usually big values enters, 
the experimental uncertainties appear generally large relative to
the obtained absolute values. Previous
experimental studies from near threshold up to 1 GeV incident energy
give a large scatter of values with a tendency of being close to zero at low
energies and increasing to values in the range of 1 - 2 mb
\cite{Rappenecker,Thomas,Gatchina,Gatchina1} towards 1 GeV, in
Ref. \cite{Dakhno} even up to 4 mb.

In Ref.~\cite{Gatchina1} the isoscalar cross sections have not been derived by
use of eq. (1). Instead of using total cross sections a partial-wave analysis
was applied to (unnormalized) angular and invariant mass
distributions. The isoscalar cross section was then extracted from the
observed asymmetries in the pion angular distribution -- -- assuming that they
exclusively derive from the interference of the isovector amplitudes with the
isoscalar ones. 

Here we report on first measurements of the isoscalar cross section from $T_p$
= 0.95 GeV up to 1.3 GeV ($\sqrt s$ = 2.3 - 2.46 GeV) by use of eq. (1). Aside
from the $\Delta(1232)$ and $N^*(1440)$ excitations, this energy range
covers the region of the $d^*(2380)$ dibaryon resonance. Whereas this
resonance is considered to decay via an intermediate $\Delta\Delta$ system in
general \cite{BR}, Kukulin and Platonova \cite{Kukulin} recently proposed an
alternative scenario, where this resonance decays into the $\Delta N$
threshold state $D_{12}$ with $I(J^P) = 1(2^+)$ by emission of a pion in
relative $p$ wave. Kinematically such a decay is hard to distinguish from
that via an intermediate $\Delta\Delta$ system. However, contrary to the
latter the decay via $D_{12}$ causes a decay branch $d^*(2380) \to
(NN\pi)_{I=0}$ because of the decay $D_{12} \to NN$. According to the SAID
partial-wave analyses \cite{SAID, SM16}, the latter decay branch is 
16 - 18$\%$. Using a total $d^*$ production cross section of about 1.7 mb
for the observed decays into $NN$ and $NN\pi\pi$ channels \cite{BR}, 
we thus expect a peak cross section of about 350 $\mu$b for the route $pn \to
d^*(2380) \to D_{12}\pi \to (NN\pi)_{I=0}$.

Following a suggestion of Bugg \cite{Bugg}, $d^*(2380)$ could represent as
well  a $N^*(1440)N$ system. Such a scenario would cause, too, a decay
of $d^*(2380)$ into the isoscalar $NN\pi$ system. Since the Roper resonance
decays into the $N\pi$ channel with a probability of 55~-~75$\%$ \cite{PDG},
we expect in this case a cross section as large as  1.1 - 1.4~mb for the route
$pn \to d^*(2380) \to N^*(1440) N \to (NN\pi)_{I=0}$.

\section{Experiment}

In order to utilize eq. (1) for the extraction of the isoscalar single-pion
production, we have measured both reactions $pp \to pp\pi^0$ and $pn \to
pp\pi^-$ simultaneously by use of their quasi-free processes in $pd$
collisions. The experiment has been carried out at COSY (Forschungszentrum
J\"ulich) at the WASA detector setup by using a proton beam with an energy of
$T_p$~=~1.2~GeV impinging on a deuterium pellet target \cite{barg,wasa}. By
exploiting the Fermi momentum of the nucleons within the deuteron in the
quasi-free scattering processes $p d \to pp\pi^0 + n_{spectator}$ and $p d \to
pp\pi^- + p_{spectator}$ , we cover the energy region $\sqrt s$ = 2.30 -
2.44 GeV (corresponding to effective incident lab energies of $T_{lab} = 0.95 -
1.3$ GeV). This includes the regions of $\Delta(1232)$, $N^*(1440)$ and
$d^*(2380)$ resonance excitations.

The hardware trigger utilized in this analysis required at least one 
charged hit in the forward detector as well as two recorded clusters in the
central detector.  

The quasi-free reaction $p d \to pp \pi^0 + n_{spectator}$
has been selected in the offline analysis by requiring one proton track in
each of the forward and central detectors as 
well as two photon hits in the central detector, which can be traced back to
the decay of a $\pi^0$ particle. The quasi-free reaction $p d \to
pp \pi^- + p_{spectator}$ has been selected in the same way with the
difference that now instead of two photon hits a $\pi^-$ track has been
required in the central detector.  
That way, the non-measured spectator four-momentum could be reconstructed by a
kinematic fit with two and one over-constraints, respectively, which derive
from the conditions for energy and momentum conservation and the $\pi^0$
mass. The achieved resolution in $\sqrt s$ was about 20 MeV.


The charged particles registered in the segmented forward detector of WASA
have been 
identified by use of the $\Delta E - E$ energy loss method. For its
application in the data analysis, all combinations of signals stemming from the
five layers of the forward range hodoscope have been used. The charged
particles in the central detector have been identified by their curved track
in the magnetic field as well as by their energy loss in the surrounding
plastic scintillator barrel and electromagnetic calorimeter.

\begin{figure} [t]
\centering
\includegraphics[width=0.8\columnwidth]{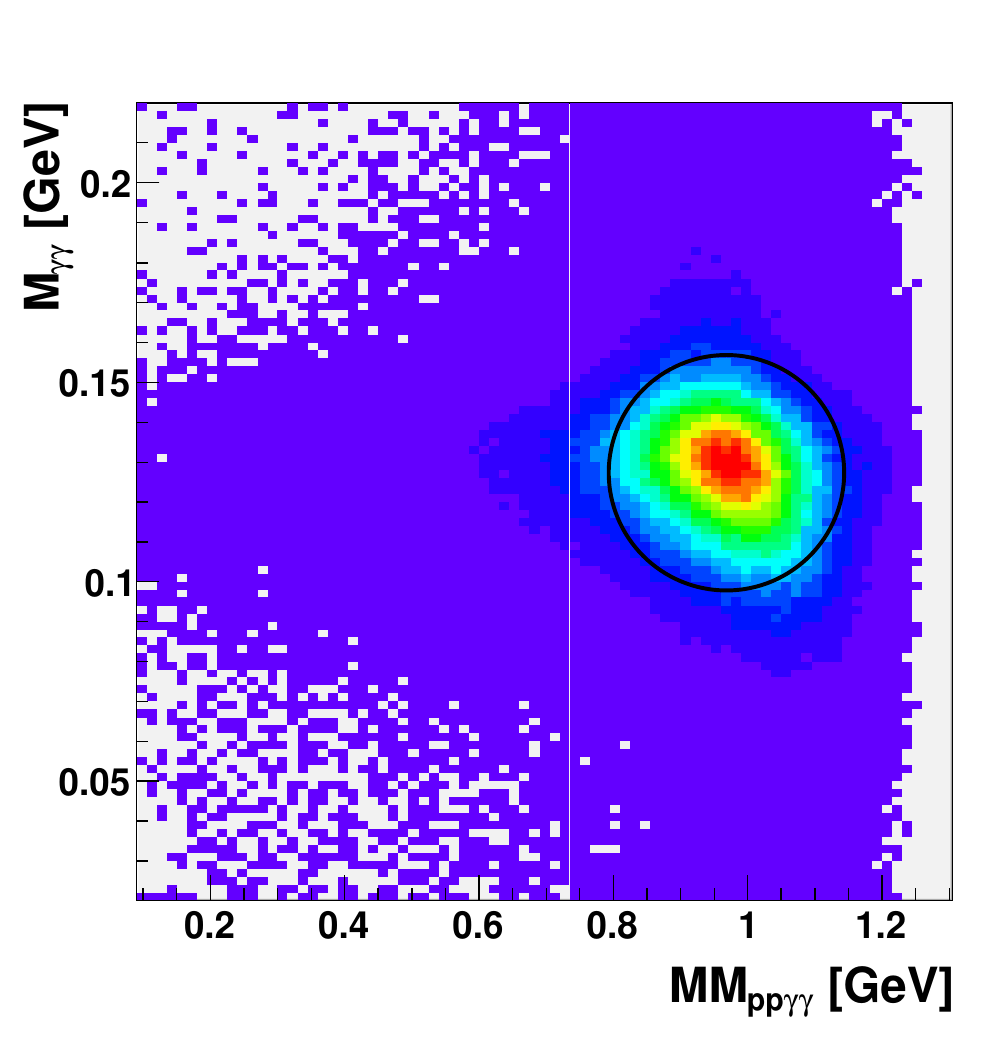}
\includegraphics[width=0.83\columnwidth]{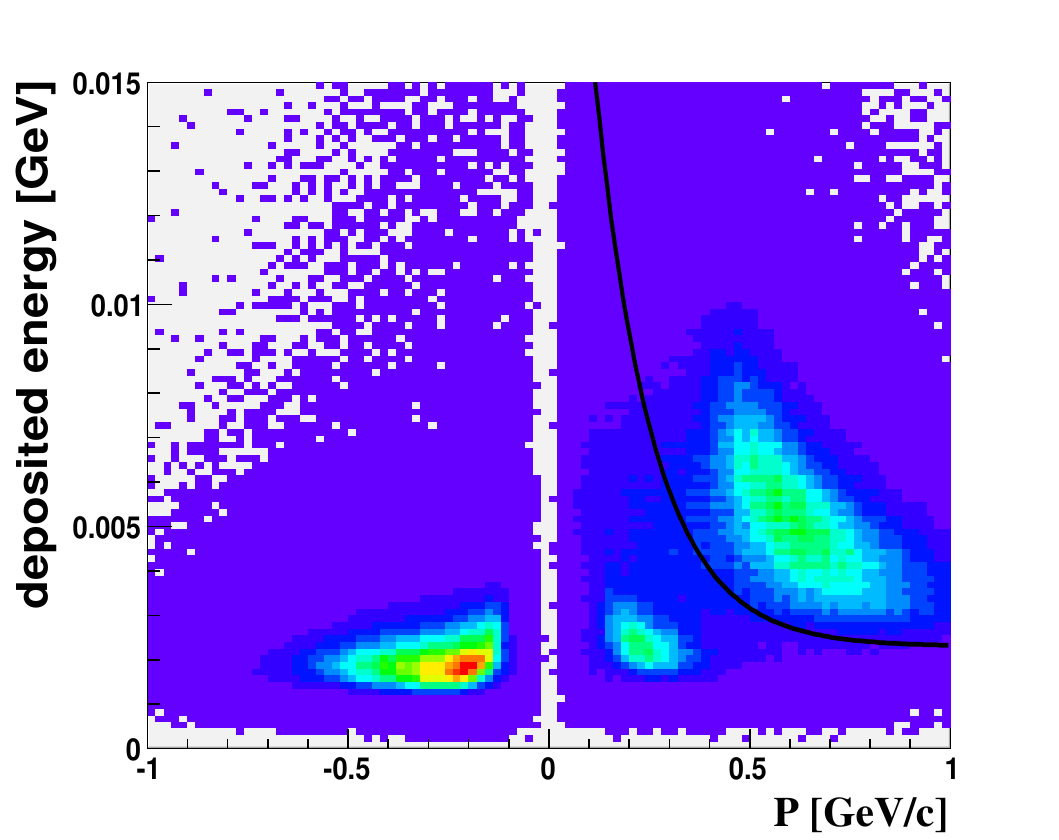}
\caption{\small (Color online) 
  Top: Plot of the $pp\gamma\gamma$-missing mass $MM_{pp\gamma\gamma}$ versus the
$\gamma\gamma$-invariant mass $M_{\gamma\gamma}$  as observed in the
quasi-free $p d \to pp \pi^0 + n_{spectator}$ reaction. The black circle denotes
the applied cut. Bottom: Plot of the momentum P of charged particles 
measured with the  minidrift chamber within the solenoid versus the
particle energy deposited in the plastic scintillator barrel for the
quasi-free  $p d \to pp \pi^- + p_{spectator}$ reaction. Negative momenta
denote negatively charged particles, positive momenta correspondingly
positively charged ones. 
}
\label{fig0}
\end{figure}

  Fig. ~\ref{fig0} shows two sample spectra to demonstrate the quality of the
  events selected for the subsequent kinematic fits. On the top panel the
  $pp\gamma\gamma$-missing mass $MM_{pp\gamma\gamma}$ is plotted versus the
  $\gamma\gamma$-invariant mass $M_{\gamma\gamma}$  as observed in the
  quasi-free $p d \to pp \pi^0 + n_{spectator}$ reaction. The black circle
  denotes the applied cut. The $\pi^0$ reconstruction efficiency is 87$\%$ for
  the case that per event 1 proton in the forward detector as well as 1
  proton and 2 neutrals in the central detector have been identified.
  On the bottom panel the momentum P of charged particles
  measured with the  minidrift chamber within the solenoid is plotted versus the
  particle energy deposited in the plastic scintillator barrel for the
  quasi-free  $p d \to pp \pi^- + p_{spectator}$ reaction. Negative momenta
  denote negatively charged particles, positive momenta correspondingly
  positively charged ones. Negative pions, positive pions and protons appear
  clearly separated. The black line shows the applied cut for the protons. The
  positive pions originate from the quasi-free two-pion production  $p d \to
  np \pi^+\pi^- + p_{spectator}$. Their leakage into the proton band causes a
  contamination of the $p d \to pp \pi^- + p_{spectator}$ reaction of less
  than 1$\%$. 

In total, a sample of about 1006800 good $pp\pi^0$ and 235000 good $pp\pi^-$
events has been selected. 
The requirement that the two protons have to be each in the angular range
covered by the forward and central detector and that the $\pi^-$ and the
gammas resulting from $\pi^0$ decay have to be in the angular range of the
central detector reduces the overall acceptance to about 38$\%$ and 41$\%$,
respectively.  The total reconstruction efficiency
including all cuts and kinematical fitting has been  
   3.0$\%$ and 0.81$\%$, respectively.
   The small efficiencies are due to the fact that we use only events,
   which have exactly three tracks in case of $pp\pi^-$ and four tracks in
   case of $pp\pi^0$, i.e., we discard events, which contain 
   beam related background like bremsstrahlung gammas.
   In addition we require the tracks to give no hint for
   secondaries due to hadronic interactions in the detector material. This is
   particularly restrictive for tracks in the forward detector due to
   high-energy protons. 

In order to understand the small total efficiencies, we illustrate the
reduction of data events by the various steps of the offline data analysis for
the case of the $pp\pi^-$ channel. In a first step all events, which meet the
trigger condition for the $pp \to pp\pi^-$ reaction, are requested to
represent charged tracks, a single one in the forward detector and
two charged tracks in the central detector. 
In the next step we require by use of the MDC information that one of
the two tracks in the central detector is of positive charge, whereas the
other one is of negative charge. This reduces the candidate events by a 
factor of four. The third step requires the three charged tracks to be
uniquely identified as two protons and one $\pi^-$ particle due to their
specific energy losses in the detector set-up without exhibiting any hadronic
interaction with the detector material. As a consequence the candidate
sample is reduced by a factor of six. The fourth step checks kinematical
conditions for emission angles and missing masses.
It removes events with corrupted kinematic information and 
leads to another factor of four reduction in the candidate events. Finally in
the last step the kinematic fit is applied, which gives another 20$\%$
reduction. 

We note that the inefficiencies of the various detector elements were studied
in rare decays studies, where {\it e.g.} in the measurement of Dalitz plot
asymmetries a 4-5 digits precision was required \cite{eta}. 

In order to check the reliability of our data analysis, comprehensive MC
simulations of reactions and detector set-up have been performed using a
cocktail of 16 reaction channels, which comprise elastic scattering,
single-pion and double-pion production. That way a realistic scenario was
generated, of how the reactions of interest were embedded in the cocktail of
background reactions. These MC simulations reproduce the event reduction in
the data  by the various analysis steps on the percent level and show that the
enrichment of proper events for the desired reaction channel by the five
analysis steps reaches 99$\%$ and better. {\it I.e.}, the contamination of the
final sample due to background reactions is at most 1$\%$.

   Efficiency and acceptance corrections of the data have been performed by MC
   simulations of reaction process and detector setup. Since the data show
   substantial deviations from pure phase-space distributions (see next
   chapter), the reaction process had to be modeled with $\Delta$ and $Roper$
   excitations, in order to achieve agreement between data and MC simulation
   and thus a consistent procedure for the data reduction process. The MC
   simulations based on pure phase-space and model descriptions will be
   discussed in the next chapter. 

Since WASA does not cover the full reaction phase space, albeit a large
fraction of it, these corrections are not fully model independent. The hatched
grey histograms in Figs. ~\ref{fig2} - ~\ref{fig5} give an estimate for these
systematic 
uncertainties. As a measure of these we have taken the difference between
model corrected results and those obtained by assuming simply pure phase
space for the acceptance corrections. 
   Though this very conservative estimate considerably exaggerates the
   systematic uncertainties, since ignoring even the well-known dominating
   $\Delta$ excitation is not physically meaningful, it nevertheless
   demonstrates the stability of the corrections.  
Compared to the uncertainties in these corrections, systematic errors
associated with modeling the reconstruction of particles are negligible.

The absolute normalization
of the data has been obtained by comparison of the 
quasi-free single pion production process $pd \to pp \pi^0 + n_{spectator}$
to previous bubble-chamber results for the $pp
\to pp \pi^0$ reaction \cite{shim,eis,hades}. That way, the uncertainty in the
absolute normalization of our data is essentially that of the previous $pp \to
pp \pi^0$ data, {\it i.e.} in the order of 5 - 15$\%$. 
For the $pn \to pp\pi^-$ reaction the extrapolation to full phase space
introduces some model dependence, which gives an uncertainty in the order of 
5 $\%$ in the absolute scale of this cross section relative to the one
of the $pp \to pp \pi^0$ reaction.

In order to have some measure of systematic uncertainties in the data
reduction process, we varied the conditions for cuts and fits within 
reasonable boundaries both in data analyses and MC simulations. As a result we
obtained a maximum of 7$\%$ change in the deduced $pn \to pp\pi^-$ cross
section relative to the one for the $pp \to pp\pi^0$ channel.


Together with the one for the acceptance correction we
end up with a systematic error of 8$\%$ for the $pp \to pp\pi^-$ cross section
relative to the one of the $pp \to pp\pi^0$ reaction.

\section{Results and Discussion}

In order to determine the energy dependence of the total cross sections for
the $pp \to pp\pi^0$ and $pn \to pp\pi^-$ reactions, we have
divided our data sample into bins of 50 MeV width in the incident energy
$T_p$. The 
resulting total cross sections for these channels as well as for the isoscalar
channel determined by use of eq. (1) are shown in Fig.~2 together with results
from earlier measurements
\cite{Dakhno,Bys,Rappenecker,Thomas,Gatchina,Gatchina1,shim,eis,hades,Dubna,brunt,Flaminio,Duna,Focardi,Baldoni,Guzhavin,Cence,bugg}. Our
data for the $pp\pi^0$ channel exhibit a flat energy dependence in good
agreement with previous data. For the $pp\pi^-$ channel our data show a slope
slightly declining with increasing energy -- also in agreement with previous
results, which exhibit some scatter.

The isoscalar cross section as obtained by use of eq. (1) is displayed at the
bottom panel of Fig.~2. Our data exhibit cross sections in the range of 2 - 5
mb. In the overlap region 
with previous results, at incident energies around 1 GeV, our data agree with
those obtained by Dakhno {\it et al.} \cite{Dakhno}, but are higher than the
results of Refs. \cite{Rappenecker,Gatchina,Gatchina1}. The latter used cross
sections for the $pp \to pp\pi^0$ reaction, which are higher by roughly 10$\%$
than those used in Ref. \cite{Dakhno}. 
   Also, in contrast to our experiment, where we measured both reactions 
   simultaneously with the same detector setup -- reducing thus systematic
   errors --, the previous results obtained at lower energies were obtained
   by use of independent measurements for $pp\pi^0$ and $pp\pi^-$ channels
   carried out 
   under different conditions, partially using even interpolations for the 
   cross section values. It is therefore not unlikely that all those
   shortcomings contribute to the large scatter of extracted values in the
   low-energy region.

By use of eq. (1) the 8 $\%$ uncertainty in the absolute scale of our cross
sections for the $pp\pi^-$ channel relative to those of the $pp\pi^0$ channel
translates to an uncertainty of 30 $\%$, {\it i.e.} about 1.5 mb, in the
absolute scale of the isoscalar cross section. Hence the discrepancy to the
results of Ref.~\cite{Gatchina1} may possibly not be as big as Fig. 2,
bottom, seems to illustrate. 

The results of Ref.~\cite{Gatchina1} agree with those of
Dakhno {\it et al.} \cite{Dakhno} for $T_p <$ 0.9 GeV. Only above there are
discrepancies. Whereas the data point from Dakhno {\it et al.}  at $T_p$ = 0.978
GeV signals a further increase of the isoscalar cross section, the cross
section deduced in Ref.~\cite{Gatchina1} starts to decrease again at higher
energies. This behavior appears to be very strange, since the Roper
excitation -- as the only isoscalar resonance process at low energies -- keeps
rising in strength 
up to 1 GeV beam energy and leveling off beyond, as we know from the analysis of
two-pion production data \cite{iso,NSTAR}. The observed energy dependence of
the isoscalar single-pion production given by the data from Dakhno
{\it et al.} \cite{Dakhno} and WASA is at least qualitatively close to that
deduced for the Roper excitation in two-pion production.

\begin{figure} 
\centering
\includegraphics[width=0.8\columnwidth]{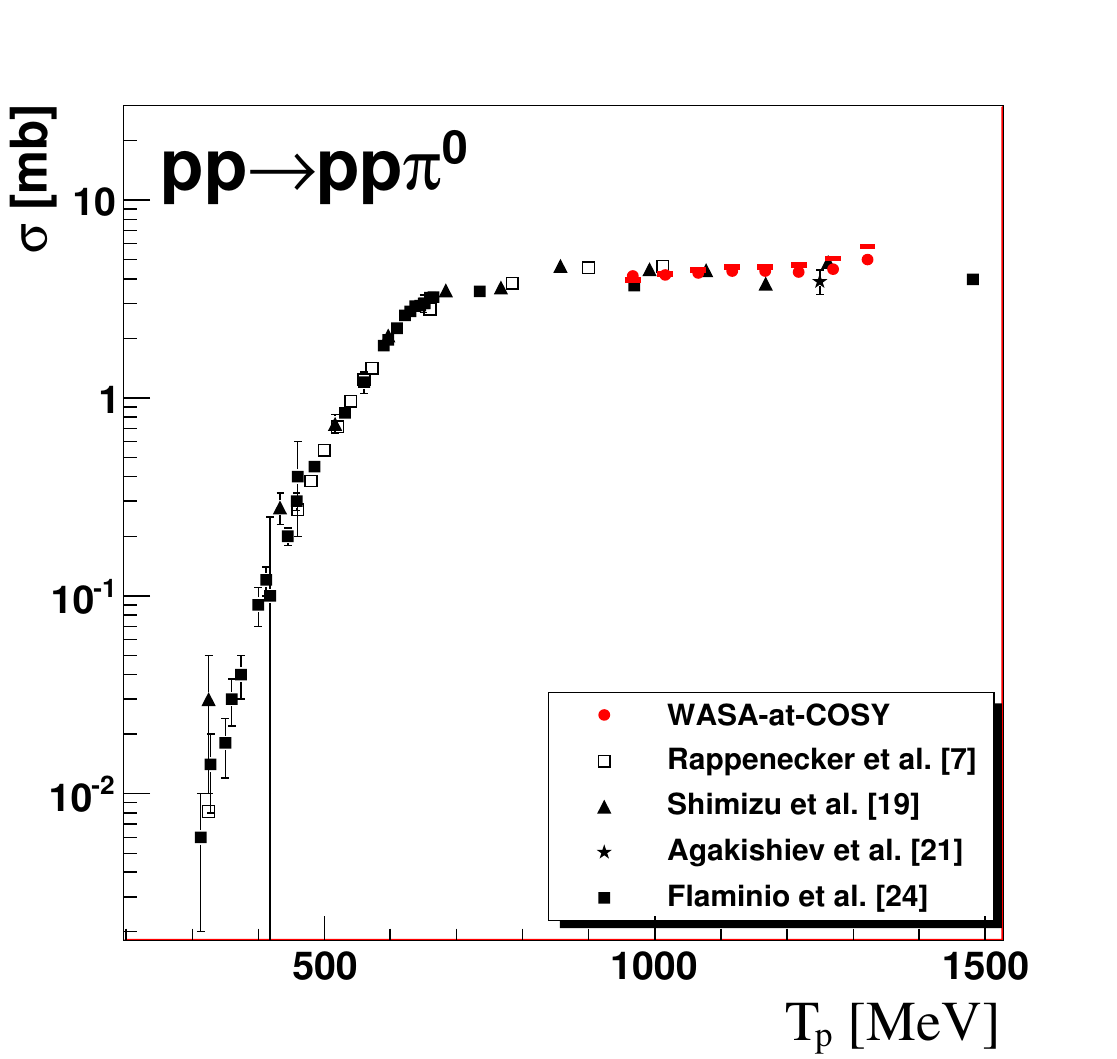}
\includegraphics[width=0.8\columnwidth]{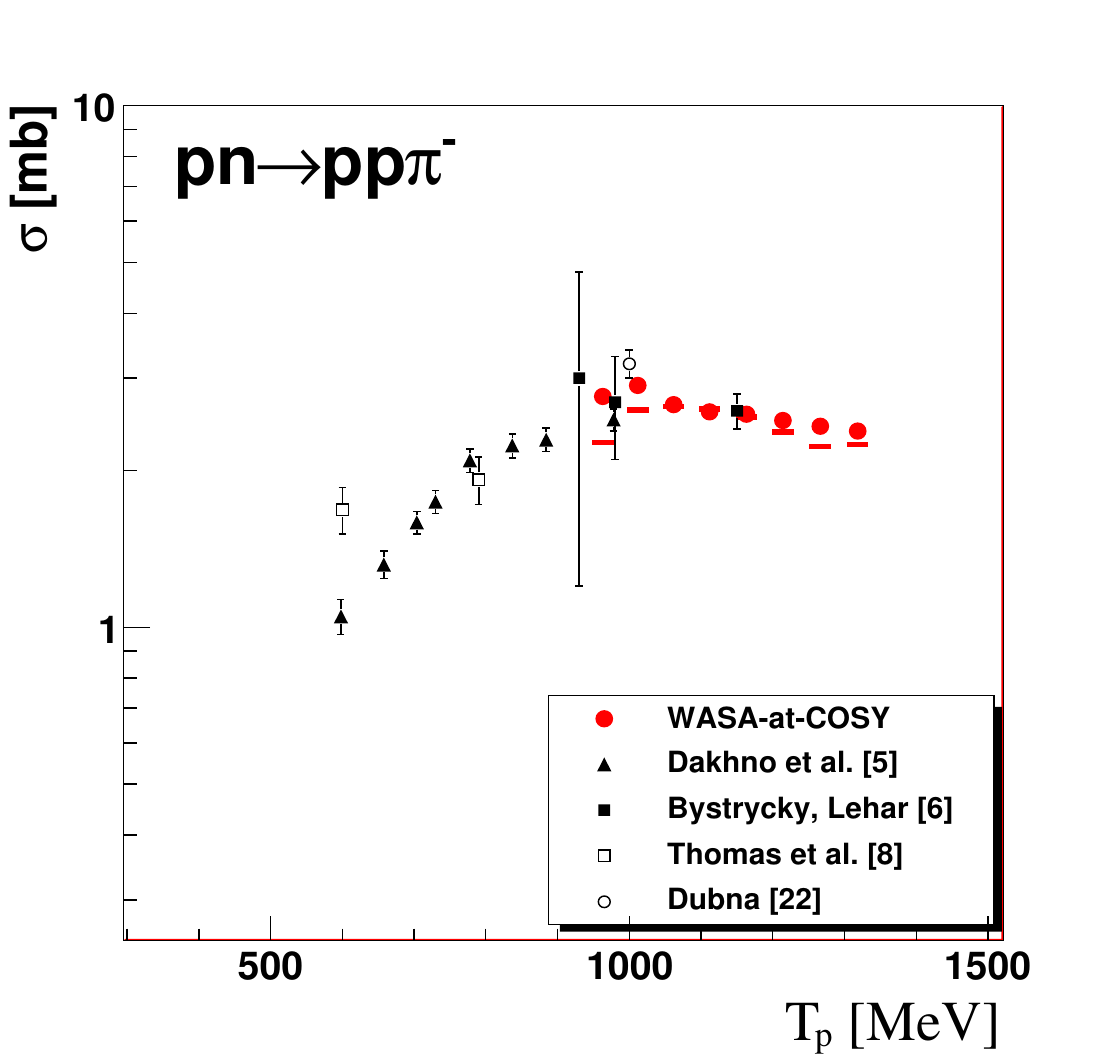}
\includegraphics[width=0.8\columnwidth]{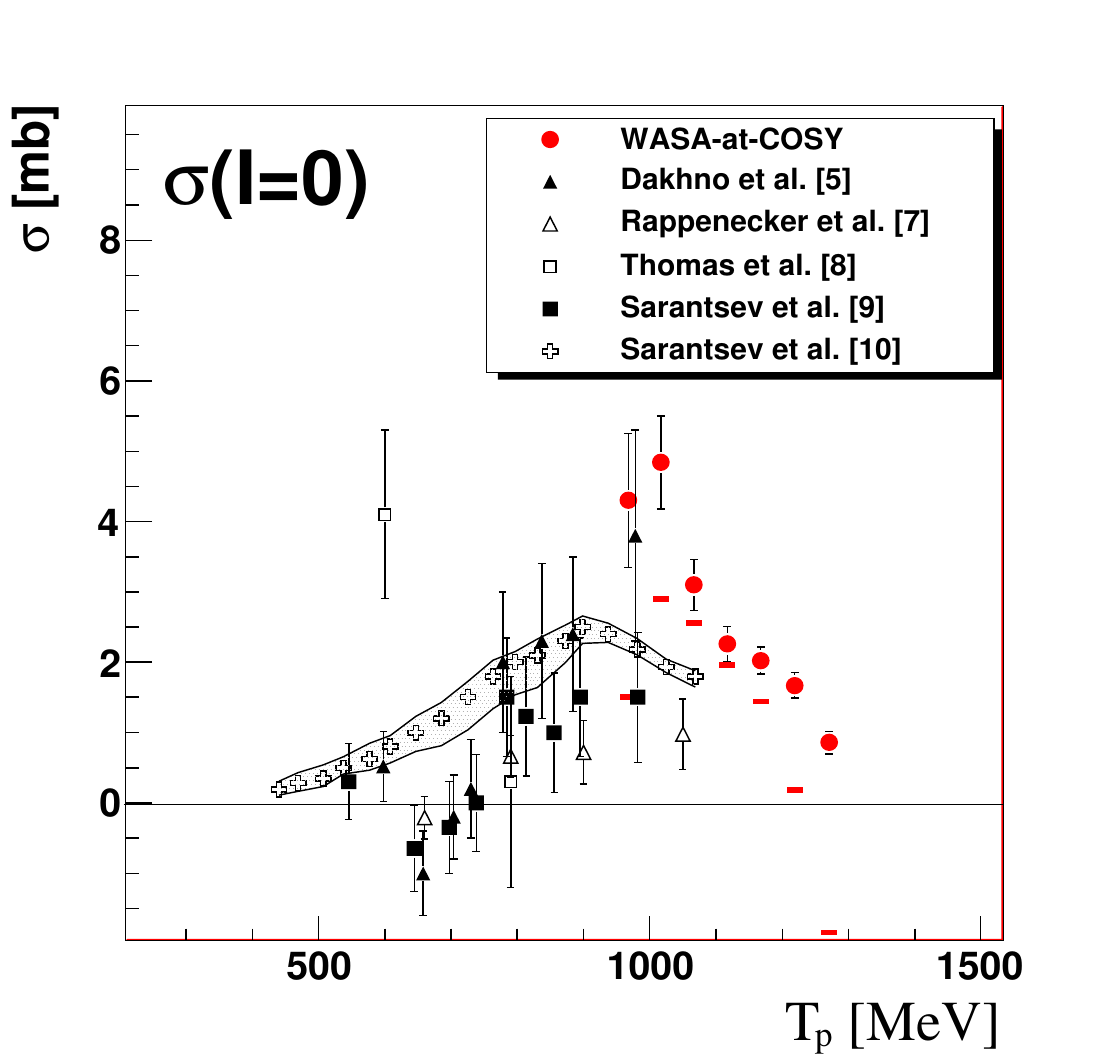}
\caption{\small (Color online) 
  Total cross sections in dependence of the incident proton energy $T_p$ for
  the reactions $pp \to pp\pi^0$ (top), $pn \to pp\pi^-$ (middle) and the
  extracted isoscalar single-pion production cross section $\sigma(I=0)$
  (bottom). Red solid circles denote the results of this work. The red
  horizontal bars represent the same data by use of a pure phase-space
  correction and serve just as an indication for 
  the stability of the results under extreme assumptions. Other symbols give
  results from earlier work \cite{Dakhno,Bys,Rappenecker,Thomas,Gatchina,Gatchina1,shim,eis,hades,Dubna,brunt,Flaminio,Duna,Focardi,Baldoni,Guzhavin,Cence,bugg}.
}
\label{fig1}
\end{figure}

\begin{figure} 
\begin{center}
\includegraphics[width=0.49\columnwidth]{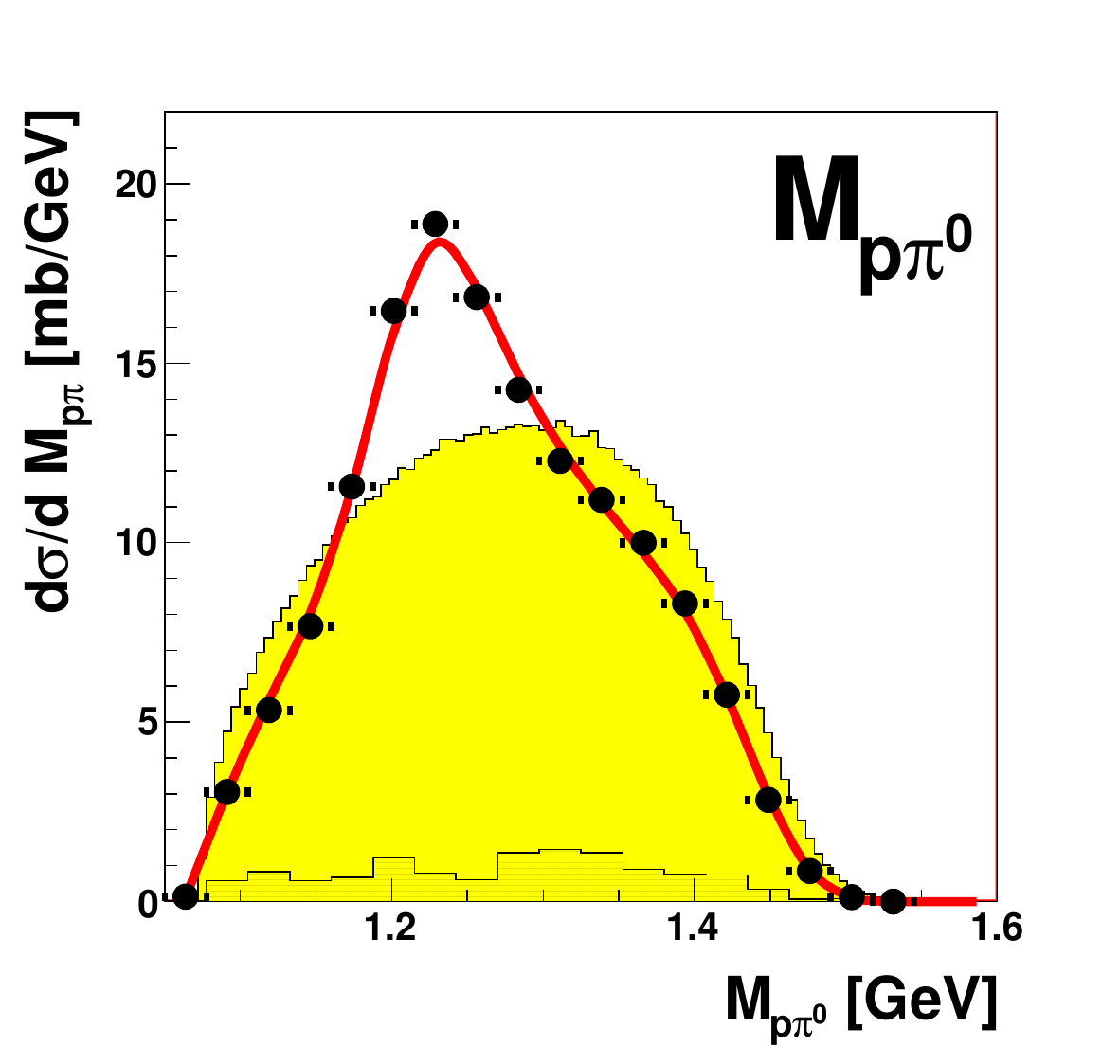}
\includegraphics[width=0.49\columnwidth]{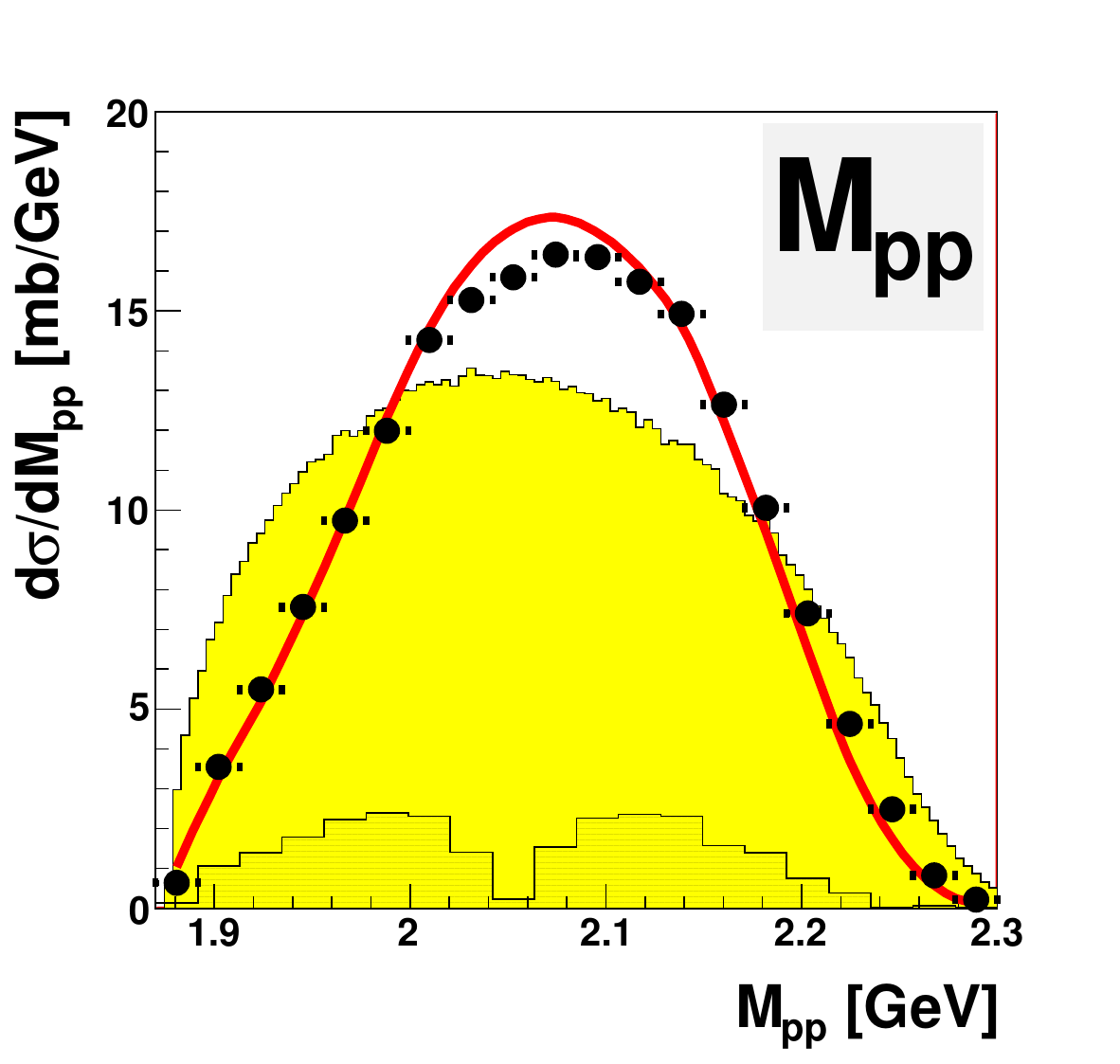}
\includegraphics[width=0.49\columnwidth]{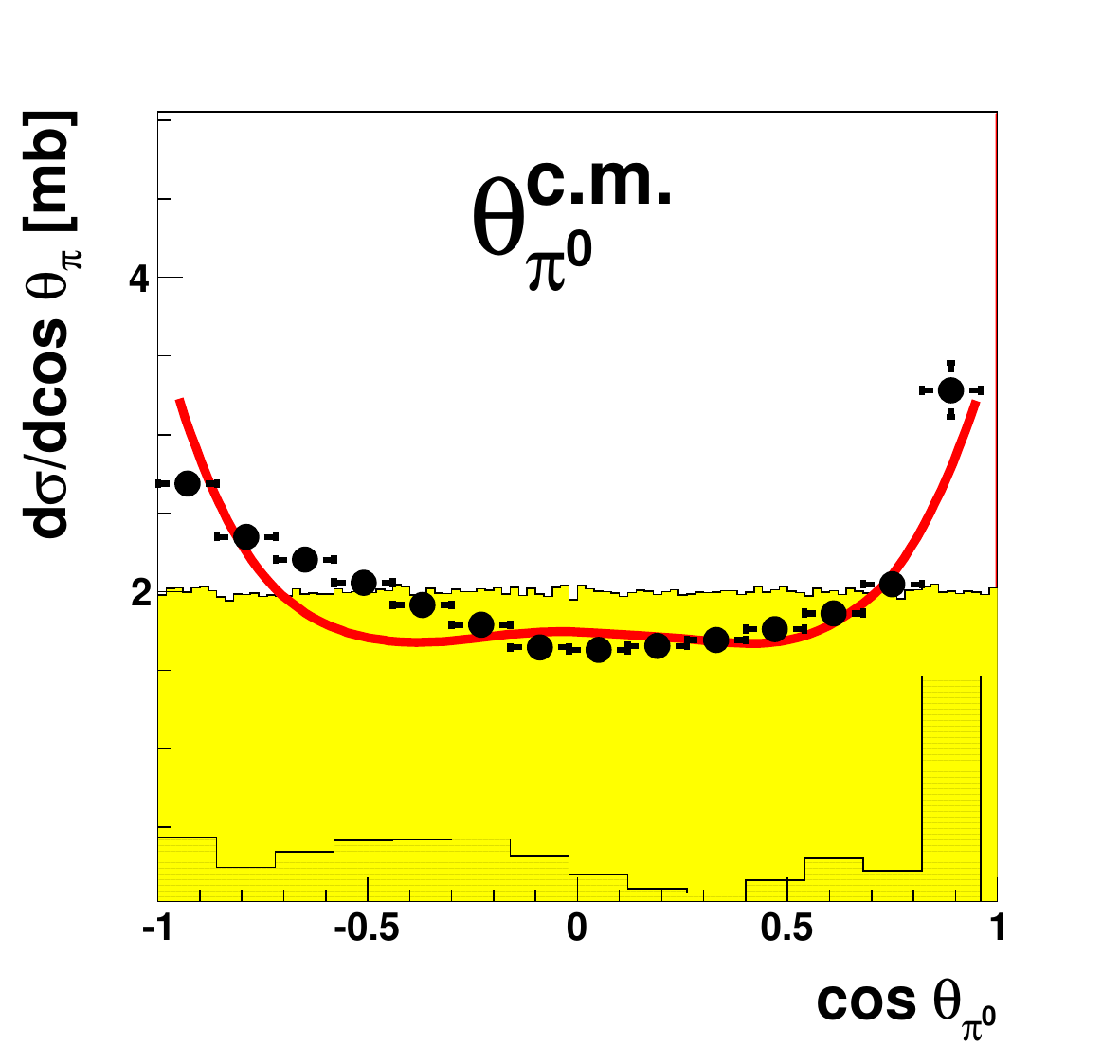}
\includegraphics[width=0.49\columnwidth]{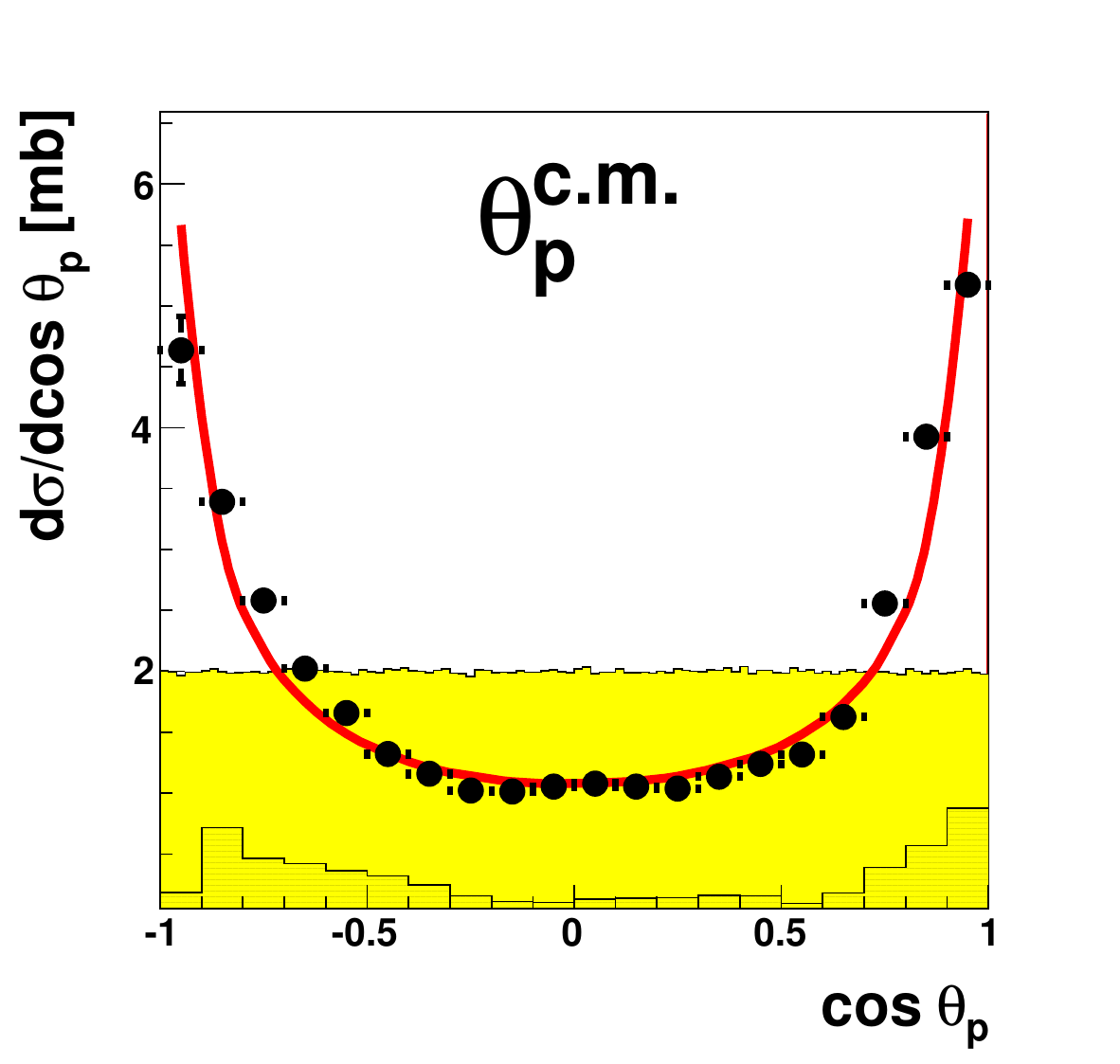}
\caption{(Color online) 
  Differential distributions of the $pp \to pp\pi^0$ reaction at $T_p$ =
  1.2. GeV for invariant-masses $M_{p\pi^0}$ (top left) and $M_{pp}$ (top right)
  of $p\pi^0$ and $pp$ subsystems, respectively, as well as  for the
  c.m. angles of neutral pions $\Theta_{\pi^0}^{c.m.}$ (bottom left) and
  protons $\Theta_p^{c.m.}$ (bottom right). The hatched histograms indicate
  systematic uncertainties due to the restricted phase-space coverage of the 
  data. The light-shaded (yellow) areas represent pure 
  phase-space distributions, the solid lines are calculations of
  $\Delta(1232)$ and $N^*(1440)$ excitations by $t$-channel meson exchange --
  normalized in area to the data.
}
\label{fig2}
\end{center}
\end{figure}

\begin{figure} 
\begin{center}
\includegraphics[width=0.49\columnwidth]{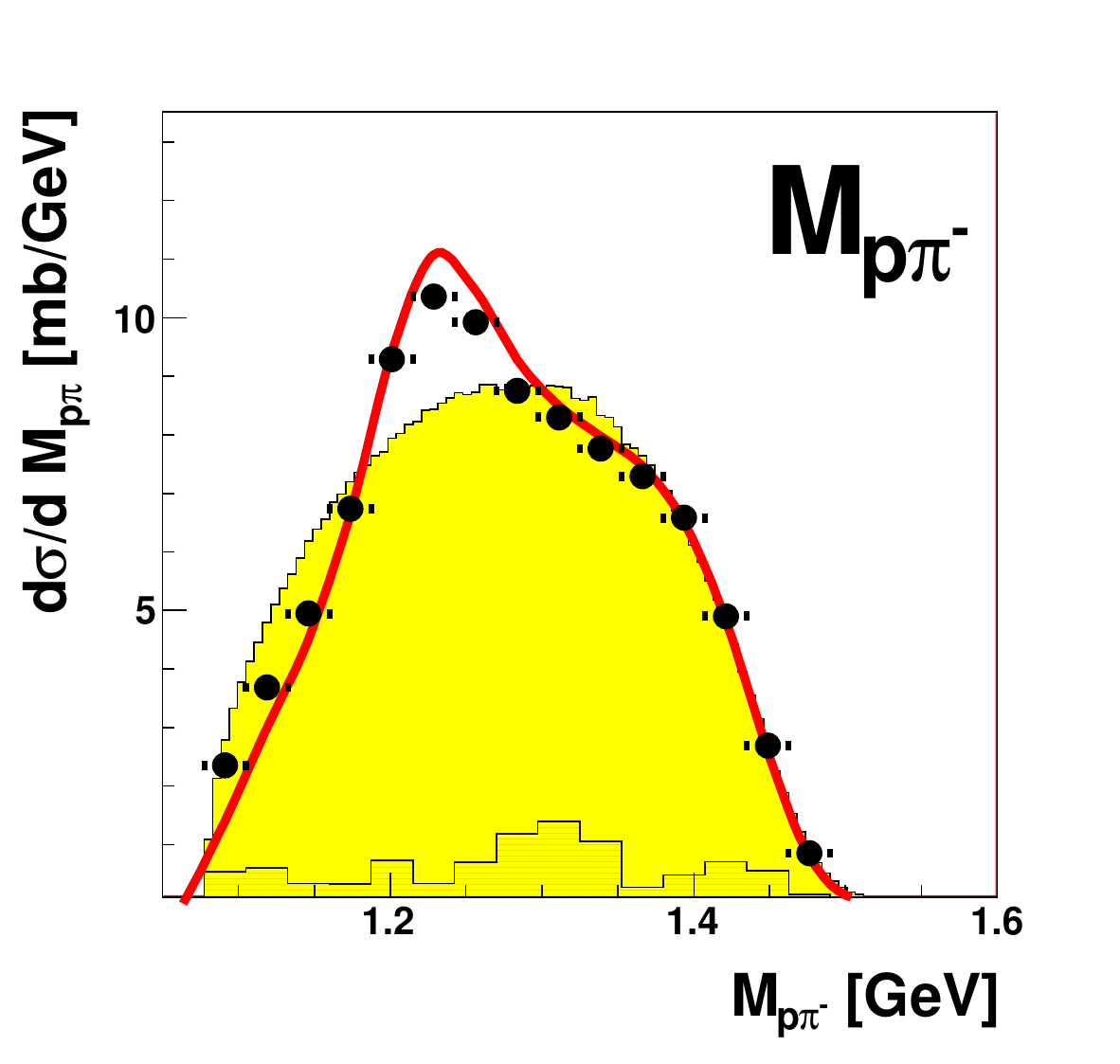}
\includegraphics[width=0.49\columnwidth]{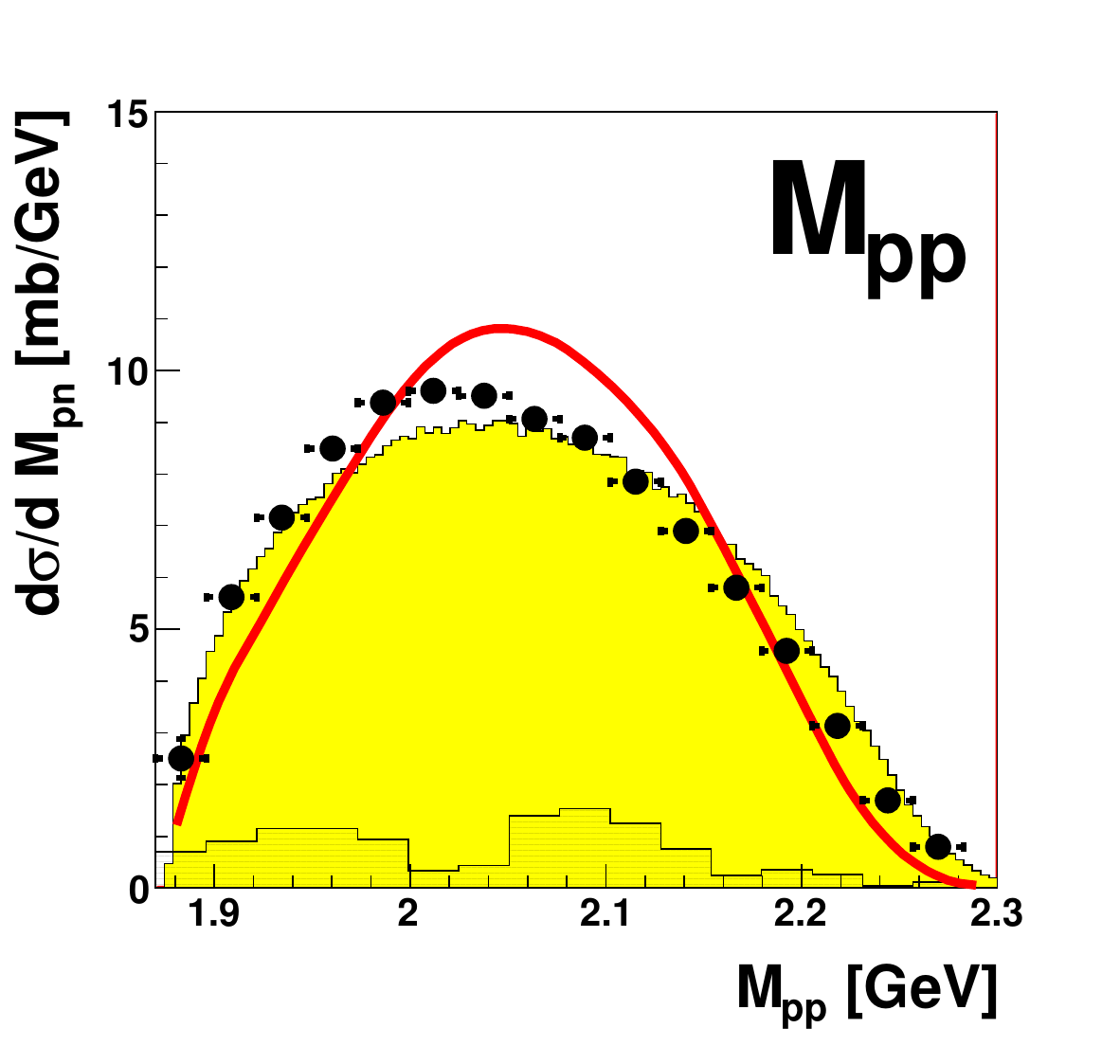}
\includegraphics[width=0.49\columnwidth]{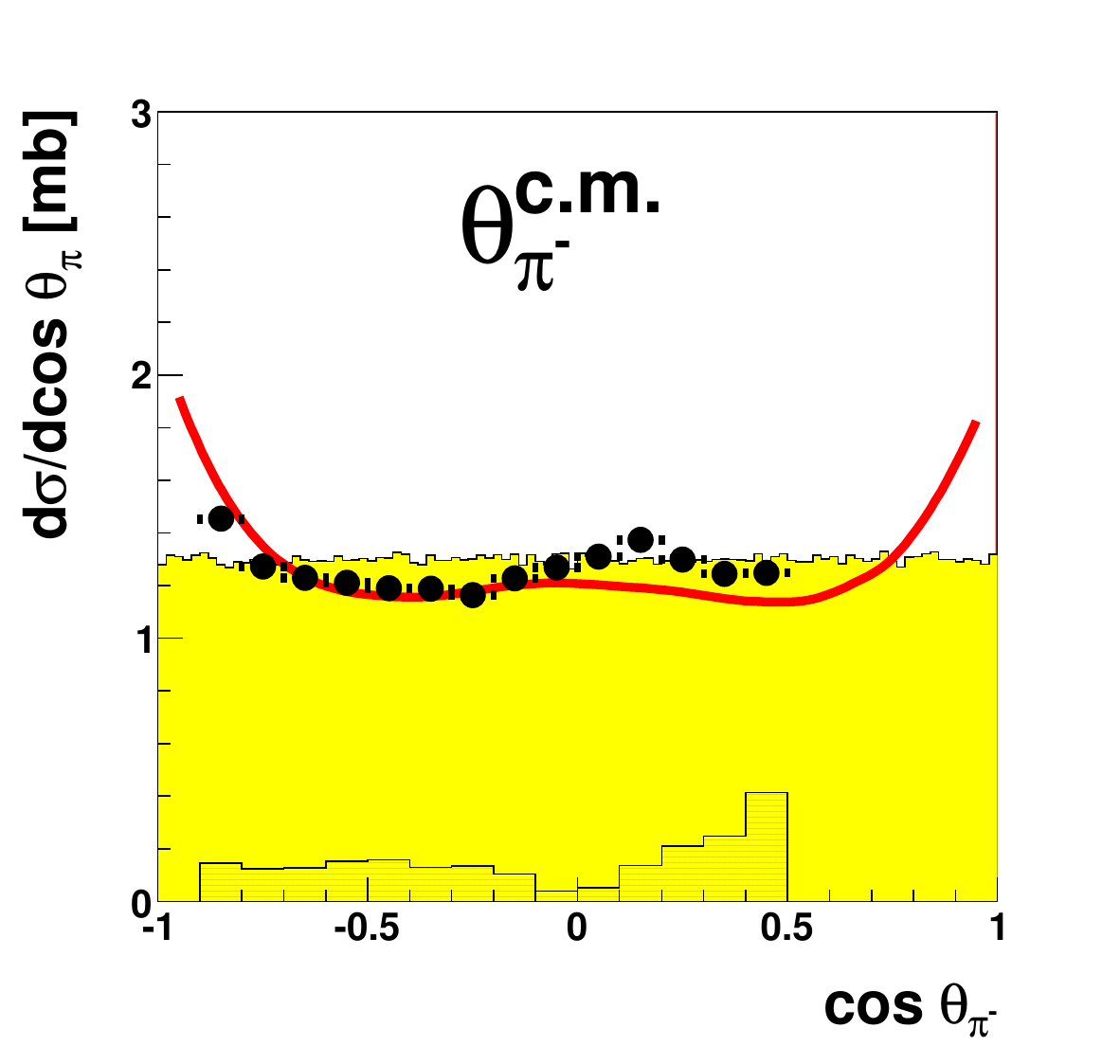}
\includegraphics[width=0.49\columnwidth]{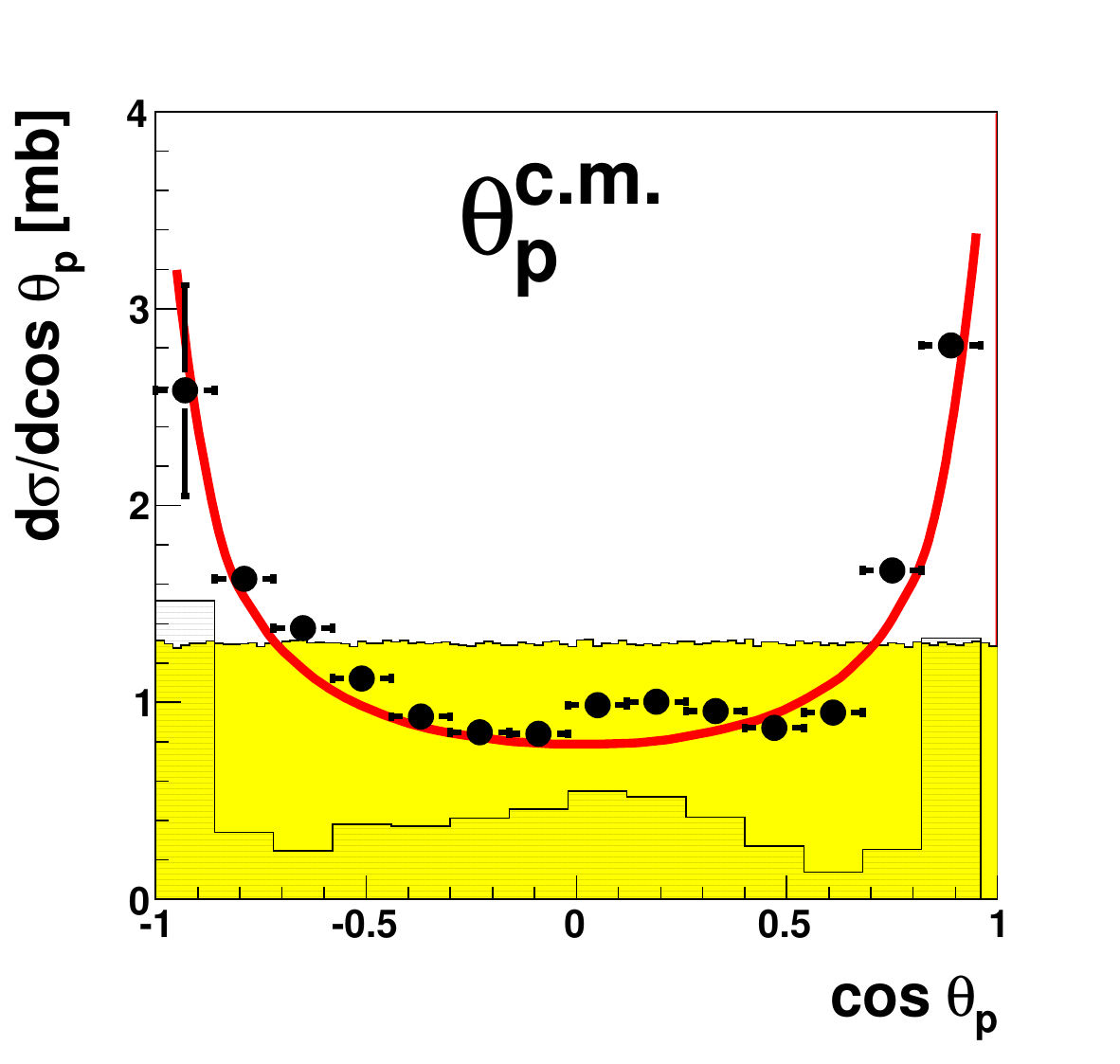}
\caption{(Color online) 
   The same as Fig. 3, but for the $pn \to pp\pi^-$ reaction.
}
\label{fig3}
\end{center}
\end{figure}

\begin{figure} [t]
\begin{center}
\includegraphics[width=0.49\columnwidth]{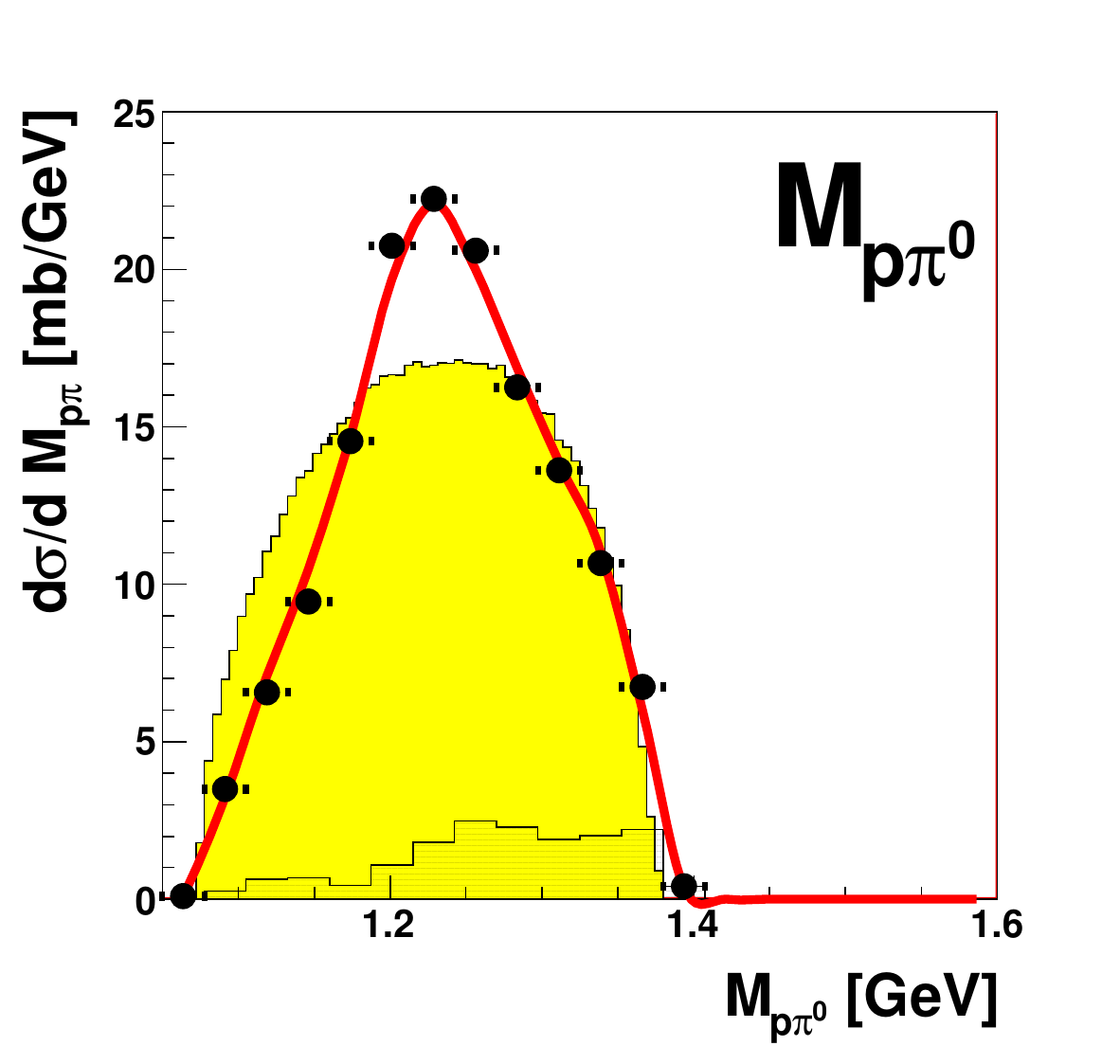}
\includegraphics[width=0.49\columnwidth]{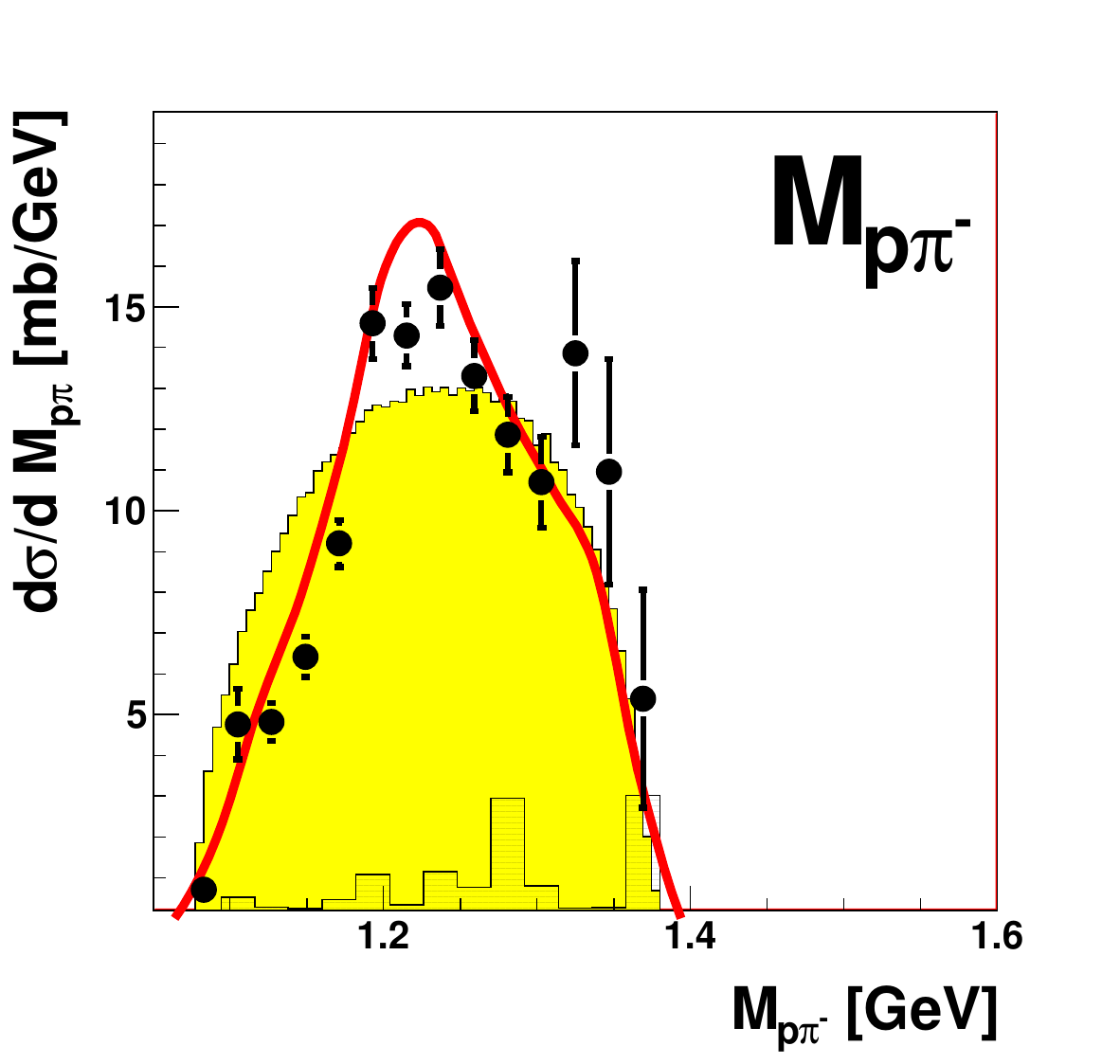}
\includegraphics[width=0.49\columnwidth]{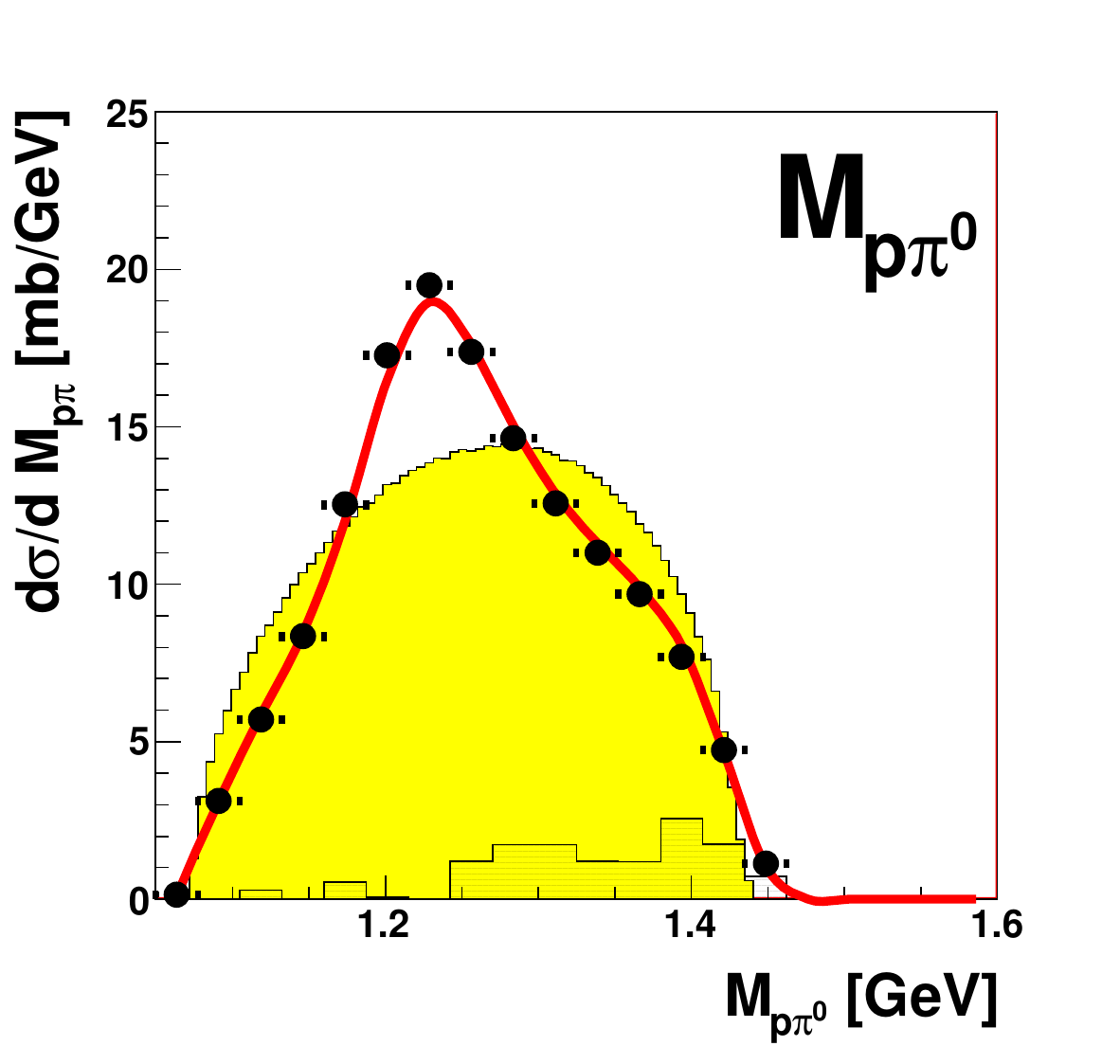}
\includegraphics[width=0.49\columnwidth]{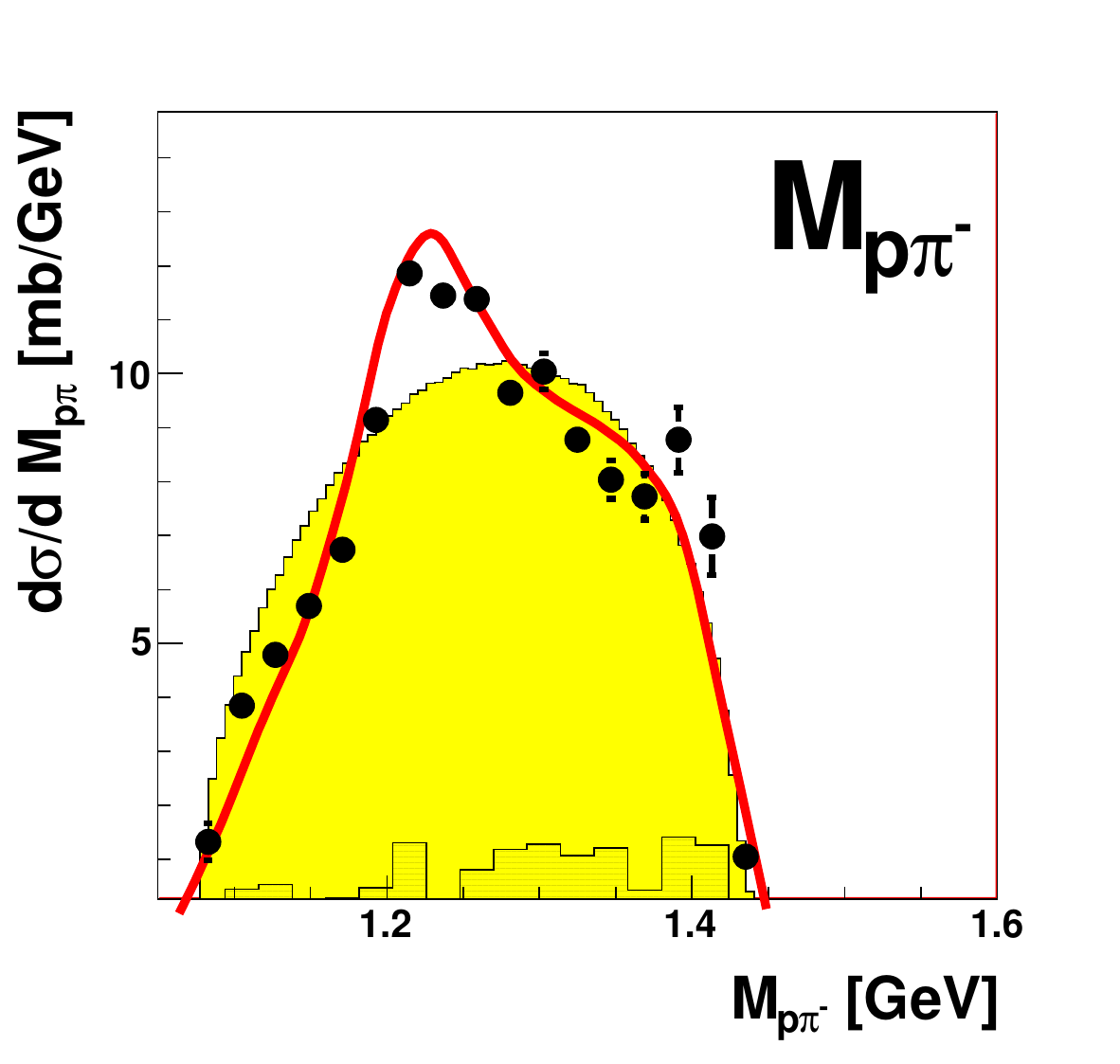}
\includegraphics[width=0.49\columnwidth]{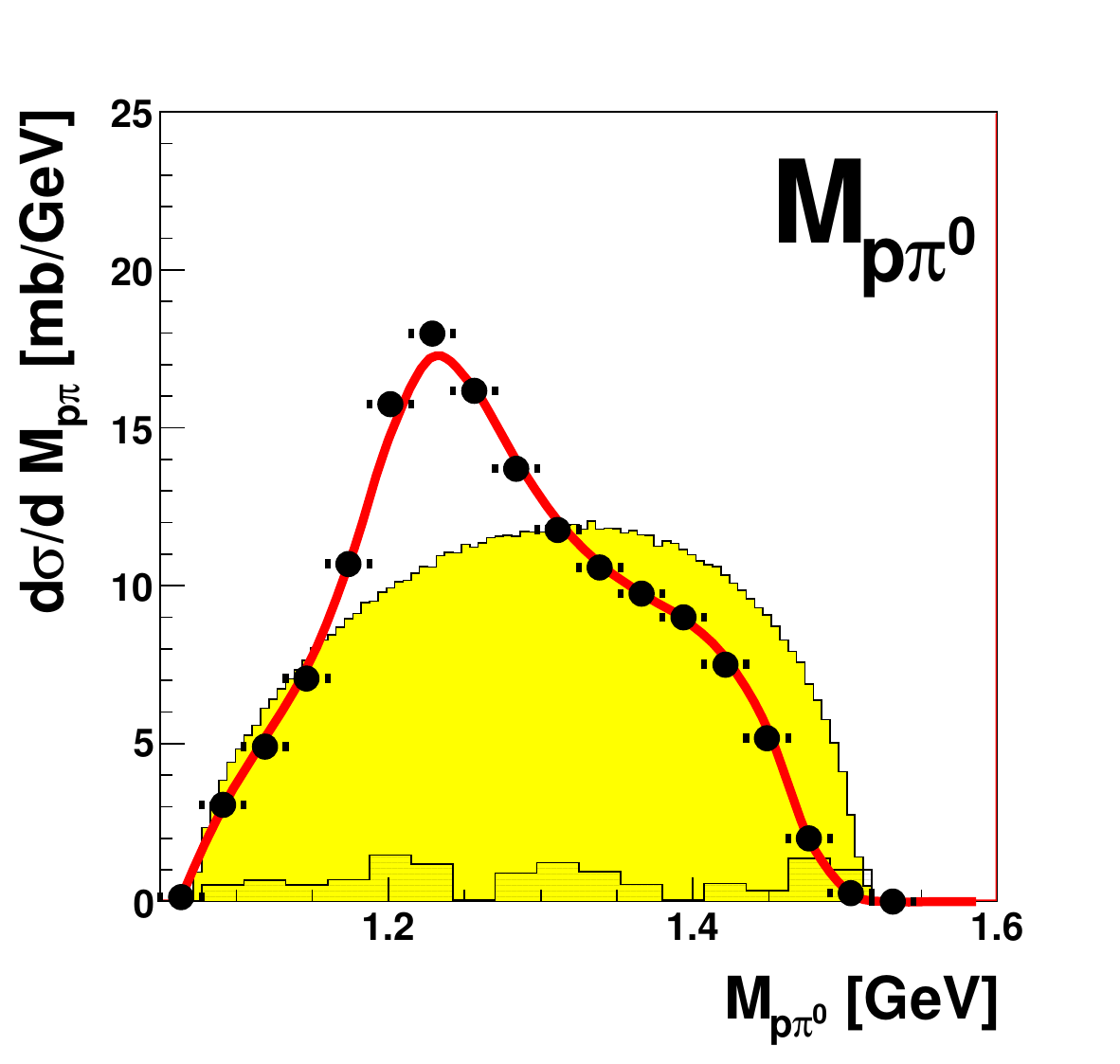}
\includegraphics[width=0.49\columnwidth]{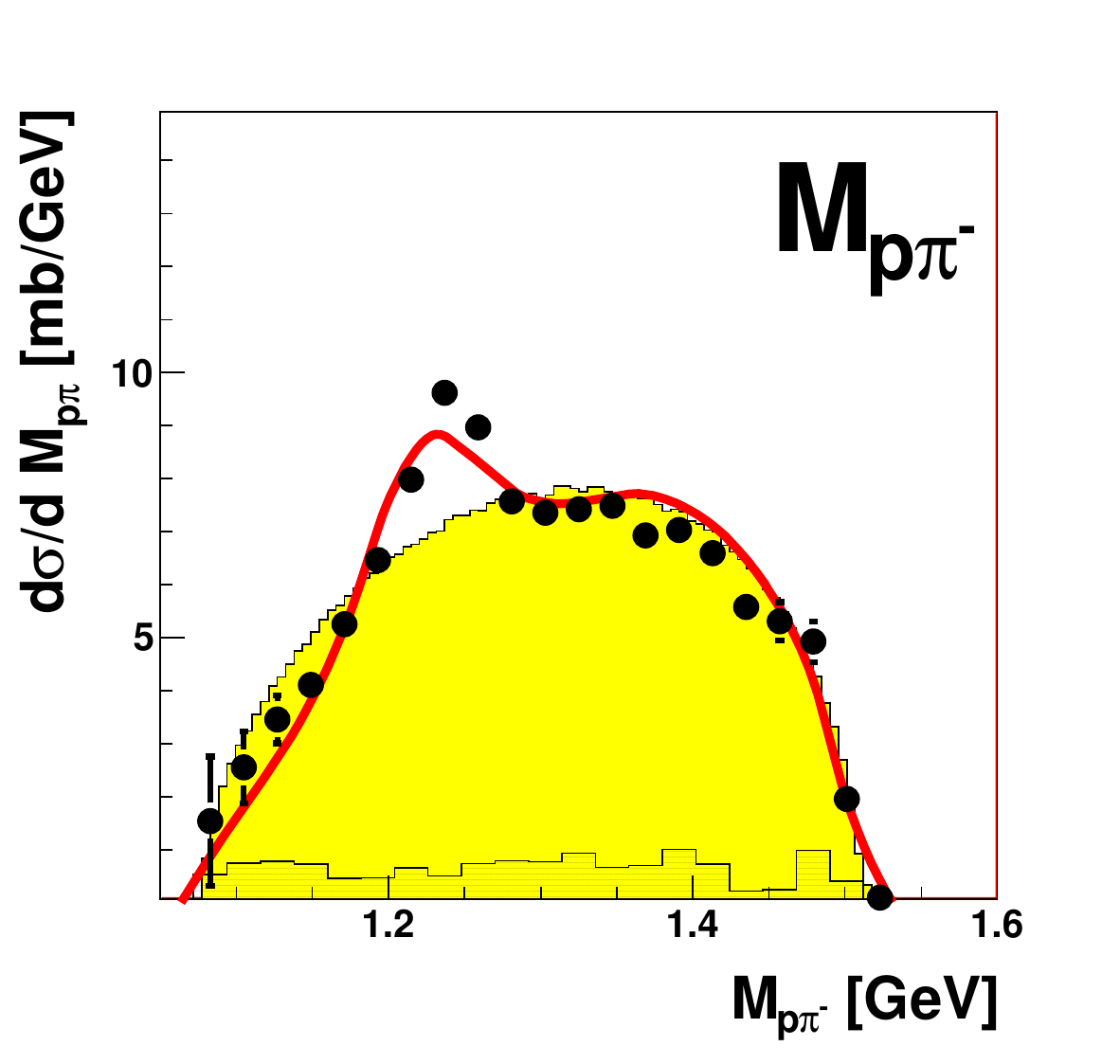}
\caption{(Color online) 
   The same as Fig. ~\ref{fig2}, but for the$M_{p\pi^0}$ (left) and $M_{p\pi^-}$
   (right) spectra for the energy bins $\sqrt s$ = 2.3 - 2.33 GeV (top), 2.34
   - 2.39 GeV (middle) and 2.40 - 2.46 GeV
   (bottom). 
}
\label{fig4}
\end{center}
\end{figure}

\begin{figure} 
\begin{center}
\includegraphics[width=0.99\columnwidth]{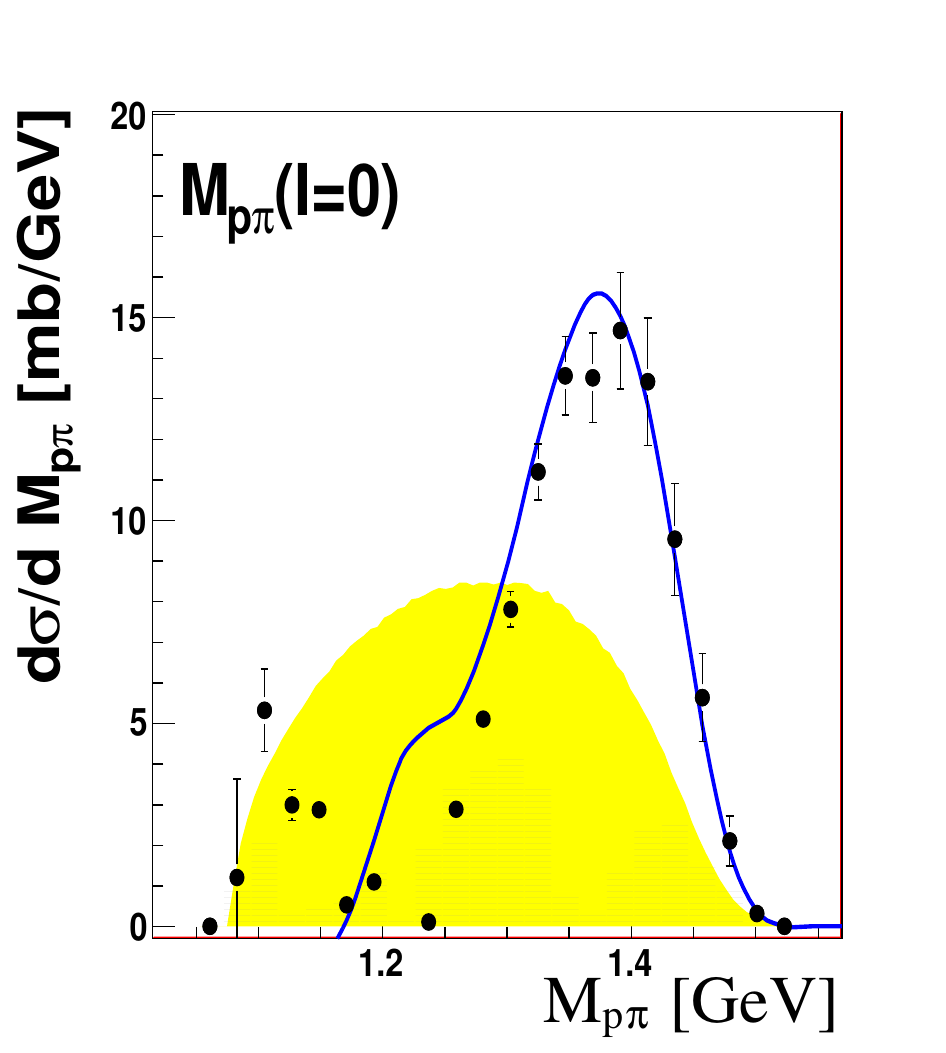}
\caption{(Color online) 
The same as Figs.~\ref{fig2} and ~\ref{fig3}, but for the the isoscalar $p\pi$
invariant mass spectrum as obtained 
  from the $M_{p\pi^0}$ and $M_{p\pi^-}$ distributions by use of eq. (1). The
  blue solid line represents the corresponding result from the $t$-channel
  calculations by use of eq. (1)  normalized in area to the data. 
}
\label{fig5}
\end{center}
\end{figure}

When binned into $\sqrt s$ bins of 20~MeV, the differential distributions do not
exhibit any particular energy dependence in their shapes -- which is of no
surprise, since the energy region covered in this measurement is dominated by
$\Delta$ and Roper excitations with very smooth energy dependencies due to
their large decay widths. Hence we
refrain from showing  
all differential distributions for single $\sqrt s$ bins. We rather
show them un-binned, {\it i.e.}, averaged over the full energy range of the
measurement, which has the advantage of better statistics and less systematic
uncertainties due to potential binning artifacts. 
  Only for the $p\pi$-invariant mass distributions we show some energy bins as
  an example.

For a three-body final state there are four independent differential
observables. We choose to show in this paper the differential distributions
for the center-of-mass (c.m.) angles for protons and pions denoted by 
$\Theta_p^{c.m.}$ and $\Theta_{\pi^0}^{c.m.}$, respectively, as well as for
the invariant masses $M_{p \pi}$ and $M_{pp}$. These distributions are shown in
Figs. ~\ref{fig2} - ~\ref{fig3}. The resolution in the angular distributions
is 0.5 - 1$^\circ$, the one in the invariant mass spectra 15~-~20~MeV. 

All measured differential distributions are markedly different in shape from
pure phase-space distributions (shaded areas in Figs. ~\ref{fig2} -
~\ref{fig5} ). They are 
reasonably well reproduced by model calculations for $t$-channel pion
exchange leading to excitation and decay of $\Delta(1232)$ and
$N^*(1440)$. This has been accomplished by utilizing the Valencia code for
pion production \cite{Luis}. The calculations are adjusted in area 
to the data in Figs. ~\ref{fig2} - ~\ref{fig5}.

The proton angular distribution is strongly forward-backward peaked in both
channels as expected for a peripheral reaction process. 
The pion angular distribution of the purely isovector $pp\pi^0$ channel, where
the $\Delta$ excitation dominates, behaves as expected from the $p$-wave
decay of the $\Delta$ resonance. For the isospin-mixed $pp\pi^-$ channel,
where the Roper resonance contributes with a flat pion angular dependence,
the observed pion angular distribution is less curved due the
combined contributions from $\Delta$ and Roper decays. The observed
asymmetries in the angular distributions are within the systematic errors.

The invariant mass spectra for $M_{p\pi^0}$ and $M_{p\pi^-}$ are both
characterized by the $\Delta$ peak -- though in the $M_{p\pi^0}$ spectrum much
more pronounced than in the $M_{p\pi^-}$ spectrum. At the high-mass shoulder of
the $\Delta$ peak the Roper excitation gets visible -- in particular now in
the $M_{p\pi^-}$ spectrum.  
  As an example for the smooth energy dependence observed in the energy interval
  of interest here, we plot the $M_{p\pi^0}$ and $M_{p\pi^-}$ spectra for
  three energy bins in Fig. ~\ref{fig4}.

By application of eq. (1) to the invariant mass spectra, we obtain the
isoscalar $p\pi$ invariant mass distribution, in which the isovector $\Delta$
process has to be absent. 
Fig.~\ref{fig5} exhibits the isoscalar $M_{p\pi}^{I=0}$
distribution, where indeed the $\Delta$ peak has vanished. The remaining
structure at higher energies has to be attributed to the isoscalar Roper
excitation (solid line). To our knowledge this is the first time that the
isoscalar Roper excitation could be visibly isolated in an invariant mass
distribution.

\begin{figure} 
\begin{center}
\includegraphics[width=0.99\columnwidth]{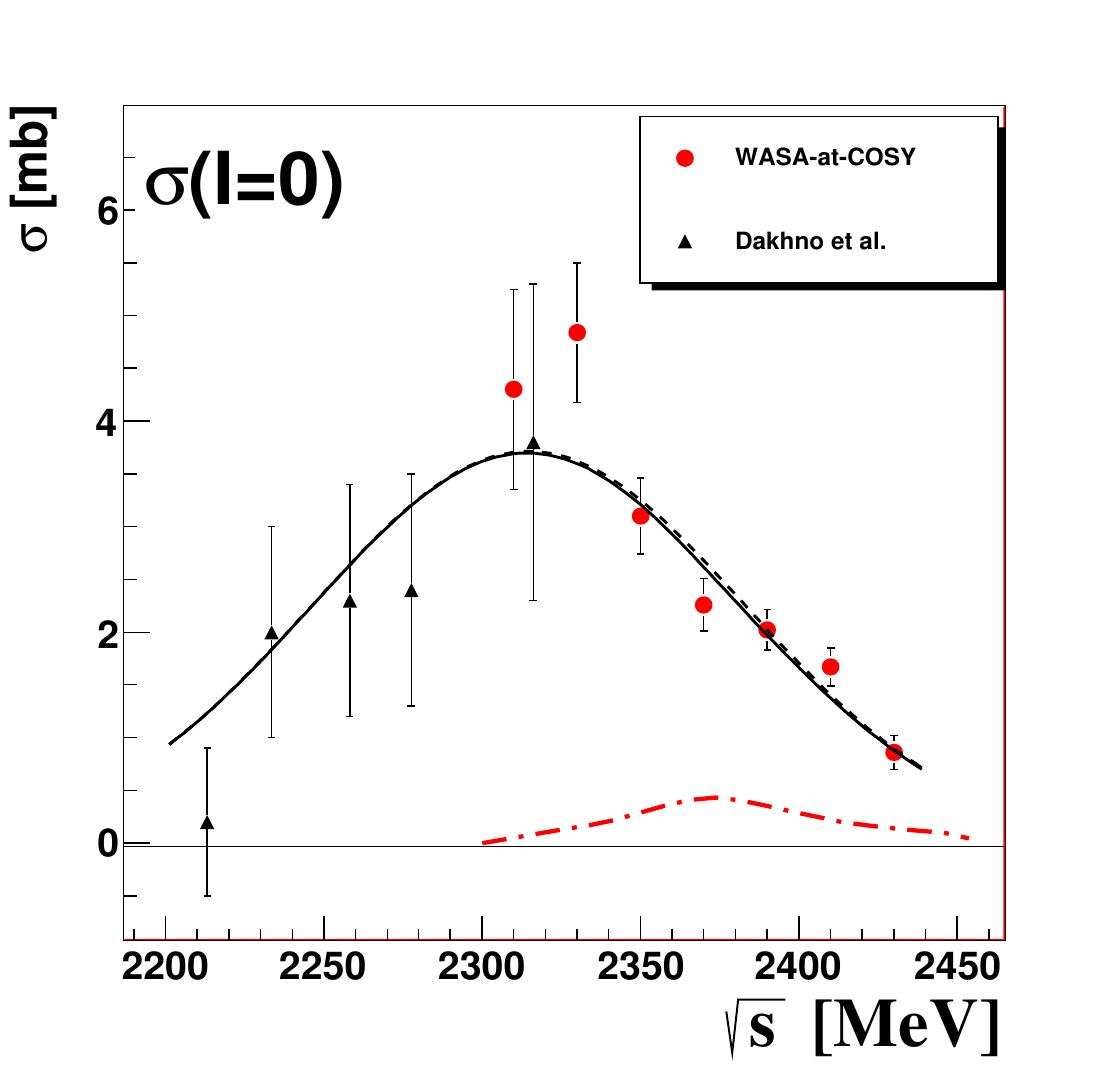}
\caption{(Color online) 
   The isoscalar single-pion production cross section in $NN$ collisions in
   dependence of the total c.m. energy $\sqrt s$. Shown are the results of
   this work (circles) together with the results from Dakhno {\it et
     al.} \cite{Dakhno} (triangles) at lower
   energies. The dash-dotted line illustrates a 10$\%$ $d^*(2380)$
   resonance contribution. Solid and dashed lines show  a fit to the
   data using a Gaussian with and without $d^*$ contribution,
   respectively. 
}
\label{fig6}
\end{center}
\end{figure}

The energy dependence of the total isoscalar cross section is displayed in
Fig.~\ref{fig6} in dependence of the c.m. energy $\sqrt s$. The only 
apparent structure is an enhancement at $\sqrt s$ = 2.33 GeV. 
However, 
its statistical significance is less than 3$\sigma$ and hence not of statistical
relevance. 

At the location of the $d^*(2380)$ resonance the cross section
exhibits no particular structure. 
In principle the resonance can interfere with the background, which is
dominated by the Roper excitation. Since we are here just 
in the region of the nominal $N^*(1440) N$ 
threshold, this system is most likely in relative $S$ wave yielding total
angular momenta of 0 and 1. In order to interfere 
with the $d^*(2380)$ resonance, the  $N^*(1440) N$ system would need to have a
total angular momentum of 3, {\it i.e.}, would need to be in relative $D$ wave
-- which is extremely unlikely at threshold. These considerations are also
supported by the partial-wave decomposition given in Ref. \cite{Gatchina1} for
the $np \to pp\pi^-$ reaction at $T_p \approx$ 1 GeV, where the contribution
of the isoscalar $^3D_3$  $np$ partial wave is below the percent level.

Since interference must be discarded, 
we have to assume that a potential $d^*(2380)$ decay into the isoscalar
$NN\pi$ channel adds incoherently to the conventional background. We observe,
however, no indication of a corresponding enhancement in the energy dependence
of the cross section. We can therefore only give an upper limit based on the statistical
uncertainty of the data points in the $d^*(2380)$ resonance region. 

The Roper
resonance can be safely assumed to produce an isoscalar contribution, which
has a very smooth energy dependence yielding a bump-like structure with a
curvature representing the large width of the Roper. A Gaussian
should therefore be a good approximation for the Roper contribution in
the region of interest. Indeed, a corresponding fit 
gives already an excellent reproduction of our data. Inclusion of a Lorentzian
representing a potential $d^*(2380)$ contribution gives an improvement only,
if a tiny negative contribution of $-35 \pm 106$ $\mu$b is allowed.  

In order to get the right curvature at the low-energy side, we next included
also the data of Dakhno {\it et al.} \cite{Dakhno} in the fit, {\it i.e.}, the
only ones, which consistently match to ours at the low-energy side. The
inclusion of these data produces a slightly stronger curvature in the region
of our data. The fit resulting in $m$ = 2310~MeV $\approx m_{Roper} + m_N$ and
$\Gamma$ = 170 MeV $\approx \Gamma_{Roper}$, which  
is shown in Fig.~\ref{fig6} by the dashed line, yields an excellent
description of both data sets. Inclusion of a potential $d^*(2380)$
contribution in the fit leads again to a negative resonance strength with a
peak value of $-65 \pm 110$ $\mu$b (solid line in  Fig.~\ref{fig6}). 
In both cases we find robust results, which are consistent with each other and
consistent with an upper limit of 180 $\mu$b at the 90~$\%$ confidence level.

However, for the decay of the dibaryon resonance $d^*(2380)$ into the isoscalar
$(NN\pi)_{I=0}$ channel via the routes $d^*(2380) \to D_{12}\pi \to (NN\pi)_{I=0}$
and $d^*(2380) \to N^*(1440)N \to (NN\pi)_{I=0}$ the relevant cross section is
not $\sigma_{NN \to NN\pi}(I=0)$, but $\sigma_{pn \to NN\pi}(I=0)$. By
definition the latter
is smaller by a factor of two \cite{Bys}. Hence the relevant value for the
upper limit for the decay branchings is 90 $\mu$b correponding    
to an upper limit for the branching ratio of  5~$\%$
for the $d^*(2380)$ decay into the $NN\pi$ channel.

This limit is much below the expectation, if $d^*(2380)$ would be dominantly a
$N^*(1440) N$ configuration. As discussed in the introduction, for this case
we expected a $d^*(2380)$ contribution of more than 1 mb. Hence this upper
limit means that only less than 7 $\%$ of the $d^*(2380)$ decay can proceed
via a $N^*(1440) N$ configuration.

For the scenario, where the $d^*(2380)$ decay proceeds via the $D_{12}\pi$
configuration, our derived upper limit is not as stringent and restricts such
a configuration only to less than 25 $\%$. Our result might be still  
compatible with a recent proposal to consider $d^*(2380)$ as a compact
hexaquark configuration surrounded by a molecule-like $D_{12}\pi$ configuration
\cite{Avraham}. It is, of course, also compatible with a pure hexaquark
scenario, where the predicted $d^* \to NN\pi$ decay rate is as small as 1 -
3$\%$ \cite{Yubing}.

\section{Conclusions}
 
The isoscalar single-pion production in $NN$ collisions has been extracted
from simultaneous measurements of the $pp \to pp\pi^0$ and $pn \to pp\pi^-$
reactions in the energy range $T_p$ = 0.95 - 1.3 GeV ($\sqrt s$ = 2.3 - 2.46
GeV). The obtained isoscalar cross sections in the region of 3 - 4 mb -- the
first ones in this energy range -- fit well to
earlier Gatchina results \cite{Dakhno} at lower energies, but less to more
recent ones \cite{Rappenecker,Gatchina}.

The differential distributions of the  $pp \to pp\pi^0$ and $pn \to pp\pi^-$
reactions are well described by $t$-channel meson exchange leading to
excitation and decay of $\Delta(1232)$ and $N^*(1440)$. 
Application of eq. (1) to invariant-mass spectra provides an isoscalar
$M_{p\pi}^{I=0}$ spectrum, where the $\Delta(1232)$ resonance is absent
leaving thus the Roper resonance isolated.

The measured energy dependence of the isoscalar cross section gives no
evidence for a decay of the dibaryon resonance $d^*(2380)$ into the
isoscalar $NN\pi$ channel. The derived upper limit excludes the proposed
$N^*(1440) N$ channel as a major intermediate decay configuration. It also
restricts a possible $D_{12}\pi$ configuration to less
than 25 $\%$, but is in full accordance with quark-model calculations
predicting a compact hexaquark configuration for $d^*(2380)$. By these
measurements the investigation of all possible hadronic decay channels of
$d^*(2380)$ has been completed. What is left, is the study of its
electromagnetic decays, which are expected to be smaller by another three to
four orders of magnitude \cite{ppnp}.

NOTE ADDED IN PROOF:
In the preceding version of this paper a factor of two was forgotten in the
derivation of the upper limits for the decay $d^*(2380) \to
(NN\pi)_{I=0}$. Hence the correct upper limits are a factor of two smaller
than given in the previous version of this article.

\section{Acknowledgments}

We acknowledge valuable discussions with A. Gal, V. Kukulin, E. Oset and
C. Wilkin on this issue. We are particularly indebted to L. Alvarez-Ruso for
using his code. 
This work has been supported by DFG (CL 214/3-1, 3-2 and 3-3) and STFC
(ST/L00478X/1).


\begin{thebibliography}{9}

\bibitem{MB} M. Bashkanov \emph{et. al.},  Phys. Rev. Lett. {\bf 102} (2009)
  052301.
\bibitem{prl2011} P. Adlarson \emph{et. al.} Phys. Rev. Lett \textbf{106} (2011)
  242302.
\bibitem{np} P. Adlarson et al., Phys. Rev. Lett \textbf{112} (2014)
  202301.
\bibitem{npfull} P. Adlarson et al., Phys. Rev. C {\bf 90} (2014) 035204.

\bibitem{Dakhno} L. G. Dakhno \emph{et. al.}, Phys. Lett. B {\bf 114} (1982)
  409. 
\bibitem{Bys} J. Bystricky \emph{et. al.}, J. Physique {\bf 48} (1987) 1901
  and references therein.

\bibitem{Rappenecker} G. Rappenecker \emph{et. al.}, Nucl. Phys. A {\bf 590}
  (1995) 763 and references therein.
\bibitem{Thomas} W. Thomas \emph{et. al.}, Phys. Rev. D {\bf 24} (1981) 1736.
\bibitem{Gatchina} V. V. Sarentsev \emph{et. al.}, Eur. Phys. J. A {\bf 21}
  (2004) 303. 
\bibitem{Gatchina1} V. V. Sarentsev \emph{et. al.}, Eur. Phys. J. A {\bf 43}
  (2010) 11.


\bibitem{BR} M. Bashkanov, H. Clement and T. Skorodko, Eur. Phys. J. {\bf 51}
  (2015) 87 and references therein.
\bibitem{Kukulin} M. Platonova and V. Kukulin, Phys. Rev. C {\bf 87} (2013)
  025202. 
\bibitem{SAID} R. A. Arndt, J. S. Hyslop III and L. D. Roper, Phys. Rev. D
  {\bf 35} (1987) 128.
\bibitem{SM16} R. L. Workman, W. J,. Briscoe and I. I. Strakovsky,
  Phys. Rev. C {\bf 94} (2016) 065203.
\bibitem{Bugg} D. V. Bugg, Eur. Phys. J. A {\bf 50} (2014) 104.
\bibitem{PDG} K. A. Olive \emph{et. al.} (PDG), Chin. Phys. C {\bf 38} (2014)
  090001.

\bibitem{barg} Ch. Bargholtz \emph{et. al.}, Nucl. Instrum. Methods A
  \textbf{547} (2005) 294.
\bibitem{wasa} H.~H.~Adam \emph{et. al.}, arXiv:nucl-ex/0411038 (2004).
\bibitem{eta} P Adlarson \emph{et. al.}, Phys. Rev. C {\bf 90} (2014) 045207.

\bibitem{shim} F. Shimizu \emph{et. al.}, Nucl. Phys. A {\bf 386} (1982) 571.
\bibitem{eis} A. M. Eisner \emph{et. al.}, Phys. Rev. {\bf 138} (1965) B670.
\bibitem{hades} G. Agakishiev \emph{et. al.}, Eur. Phys. J. A {\bf 51} (2015)
  137. 

\bibitem{Dubna} A. Abdivaliev \emph{et. al.}, Dubna Preprint JINR D1-81-756
  (1981).
\bibitem{brunt} D. C. Brunt, M. J. Clayton and B. A. Westwood, Phys. Rev. {\bf
    187} (1969) 1856. 
\bibitem{Flaminio} V. Flaminio \emph{et. al.}, CERN libraries, CERN-HERA 84-01
  (1984).
\bibitem{Duna} A. F. Dunaytsev and Y. D. Prokoshkin, Sov. Phys. JETP {\bf 9}
  (1959) 1179.
\bibitem{Focardi} S. Focardi\emph{et. al.}, Nuovo Cim. {\bf 39} (1965) 405.
\bibitem{Baldoni} B. Baldoni \emph{et. al.}, Nuovo Cim. {\bf 26} (1962) 1376. 
\bibitem{Guzhavin} V. M. Guzhavin  \emph{et. al.},  Sov. Phys. JETP {\bf 19}
  (1964) 847.
\bibitem{Cence} R. J. Cence  \emph{et. al.}, Phys. Rev. {\bf 131} (1963) 2713.
\bibitem{bugg} D. V. Bugg  \emph{et. al.}, Phys. Rev. {\bf 133} (1964) B1017.

\bibitem{iso} T. Skorodko \emph{et. al.}, Phys. Lett. B {\bf 679} (2009) 30.
\bibitem{NSTAR} T. Skorodko \emph{et. al.}, Eur. Phys. J. A {\bf 35} (2008) 317.

\bibitem{Luis} L. Alvarez-Ruso, E. Oset, E. Hernandez, Nucl. Phys. A {\bf
    633} (1998) 519 and priv. comm. 
\bibitem{Avraham} A. Gal, Phys. Lett. B {\bf 769} (2017) 436.
\bibitem{Yubing} Yubing Dong, Fei Huang, Pengnian Shen and Zongye Zhang,
  Phys. Lett. B {\bf 769} (2017) 223.
\bibitem{ppnp} H. Clement, Prog. Part. Nucl. Phys. {\bf 93} (2017) 195.
 
\end{thebibliography}
\end{document}